# A triple star system with a misaligned and warped circumstellar disk shaped by disk tearing

Stefan Kraus,[1]* Alexander Kreplin,[1] Alison K. Young,[1,2] Matthew R. Bate,[1]
John D. Monnier,[3] Tim J. Harries,[1] Henning Avenhaus, Jacques Kluska,[1,4]
Anna S. E. Laws,[1] Evan A. Rich,[3] Matthew Willson,[1,5] Alicia N. Aarnio,[6]
Fred C. Adams,[3] Sean M. Andrews,[7] Narsireddy Anugu,[1,3,8] Jaehan Bae,[3,9]
Theo ten Brummelaar,[10] Nuria Calvet,[3] Michel Curé,[11] Claire L. Davies,[1]
Jacob Ennis,[3] Catherine Espaillat,[12] Tyler Gardner,[3] Lee Hartmann,[3]
Sasha Hinkley,[1] Aaron Labdon,[1] Cyprien Lanthermann,[4]
Jean-Baptiste LeBouquin,[3,13] Gail H. Schaefer,[10] Benjamin R. Setterholm,[3]
David Wilner,[7] Zhaohuan Zhu[14]

[1]University of Exeter, School of Physics & Astronomy, Exeter, EX4 4QL, UK,
[2]School of Physics and Astronomy, University of Leicester, Leicester, LE1 7RH, UK,
[3]University of Michigan, Ann Arbor, MI 48109, USA,
[4]Instituut voor Sterrenkunde, Katholieke Universiteit Leuven, 3001 Leuven, Belgium,
[5]Georgia State University, Atlanta, GA 30302, USA,
[6]University of North Carolina Greensboro, Greensboro, NC 27402, USA,
[7]Center for Astrophysics — Harvard & Smithsonian, Cambridge, MA 02138, USA,
[8]Steward Observatory, University of Arizona, Tucson, AZ 85721, USA,
[9]Carnegie Institution for Science, Washington, DC 20015, USA,
[10]The Center for High Angular Resolution Astronomy Array
of Georgia State University, Mount Wilson, CA 91023, USA,
[11]Instituto de Fisica y Astronomia, Universidad de Valparaiso, Casilla 5030, Valparaiso, Chile
[12]Boston University, Boston, MA 02215, USA,
[13]Université Grenoble Alpes, Institut de Planétologie et d'Astrophysique, 38000 Grenoble, France,
[14]University of Nevada, Las Vegas, NV 89154, USA

*To whom correspondence should be addressed; E-mail: s.kraus@exeter.ac.uk

**Young stars are surrounded by a circumstellar disk of gas and dust, within which planet formation can occur. Gravitational forces in multiple star systems can disrupt the disk. Theoretical models predict that if the disk is misaligned with the orbital plane of the stars, the disk should warp and break into precessing rings, a phenomenon known as disk tearing. We present observations of the triple star system GW Orionis, finding evidence for disk tearing. Our images show an eccentric ring that is misaligned with the orbital planes and the outer disk. The ring casts shadows on a strongly warped intermediate region of the disk. If planets can form within the warped disk, disk tearing could provide a mechanism for forming wide-separation planets on oblique orbits.**

Stars form through fragmentation & collapse of molecular clouds. The most frequent outcome of this process is a gravitationally bound multiple star system, such as a binary or triple (*1, 2*). As the system evolves, the stars interact dynamically with each other and with the surrounding circum-multiple disk of gas and dust, which holds material that could either accrete onto the stars or form planets. Numerical simulations (*3, 4*) have predicted a novel hydrodynamic effect known as disk tearing in the disks around multiple systems if the orbital plane of the stars is strongly misaligned with the disk plane. Gravitational torque from the stars is predicted to break the disk into several distinct planes, forming rings. These rings should separate from the disk plane and precess around the central stars (*5*). Misaligned disks have previously been observed but it has not been possible to link the misaligned structures clearly to disk tearing, either due to the non-detection of the pertuber (e.g. (*6*)) or insufficient constraints on the orbit (e.g. (*7–9*)).

We present observations of GW Orionis, a young [$1.0 \pm 0.1$ million years old (*10*)] triple star system located in the $\lambda$ Orionis region of the Orion Molecular Cloud, whose central cluster is at a distance of $388 \pm 5$ parsec (*11*). The system consists of a close (1.2 astronomical unit, au) binary with a $\sim 242$ days period on a nearly-circular orbit (stars GW Ori A and GW Ori B; (*12, 13*)) and a tertiary that orbits in $\sim 11$ years at $\sim 8$ au separation (GW Ori C; (*14, 15*)).

We monitored the orbital motion of the system over 11 years using near-infrared interferometry (1.4-2.4 $\mu$m thermal continuum emission; Fig. S8). Fitting an orbit model to these observations (*16*) results in tight constraints on the masses of the three stars (GW Ori A: $2.47 \pm 0.33$, GW Ori B: $1.43 \pm 0.18$, and GW Ori C: $M_C = 1.36 \pm 0.28$ solar masses) and the orientation of the orbits. The orbits of the inner pair (A-B) and the tertiary (AB-C) are tilted $13.9 \pm 1.1°$ from each other.

We imaged the system using sub-millimeter and near-infrared interferometry, which trace thermal dust emission, and using visible and near-infrared adaptive-optics imaging polarimetry, which trace scattered light. These observations allow us to constrain the dust distribution in the system. Combining these techniques enables us to constrain the 3-dimensional orientations of the disk components and search for disk warping. The cold dust (down to $\sim 10$ K dust temperature, traced by 1.3mm continuum emission) is arranged in three rings. The two outer rings (with radii of $334 \pm 13$ and $182 \pm 12$ au; labeled R1 and R2 in Fig. 1A) are centered on the A-B binary and seen at inclinations of $142 \pm 1°$ and $143 \pm 1°$ from a face-on view. This corresponds to retrograde rotation (in clockwise direction on the sky) with the Eastern side tilted towards us by $38°$ and $37°$ for R1 and R2. The third, innermost ring R3 has a projected radius of $43.5 \pm 1.1$ au and appears more circular in projection than R1 and R2. R3 is offset with respect to the center-of-mass of the system (Fig. 1B). Dust emission is apparent between the rings as well as inside R3, with a factor $\sim 10$ lower flux density than in the neighboring rings.

Our infrared polarimetric images show asymmetric scattered light extending from $\sim 50$ to $\sim 500$ au. The scattered light forms four arcs A1 to A4 (Figs. 1C and 1D) with the Eastern side appearing brighter than the Western side. This is consistent with the Eastern side of the disk facing towards Earth. The dimmer regions separating the arcs A1, A2, and A3 coincide with the dust rings R1, R2, and R3 seen in the sub-millimeter image. We interpret this as a shadowing effect where the increased disk scale height at the location of dust rings R1, R2, and R3 casts a shadow on the flared disk (Fig. S3, (*16*)). We interpret arcs A3 and A4 as parts of a single elliptical structure, whose semi-major axis orientation (along position angle, PA$\sim 30°$, measured East of North) deviates from the orientation of the outer disk (which has PA$\sim 0°$). Two sharp shadows, S1 and S2, extend in radial direction. The Eastern shadow S1 changes direction at $\sim 100$ au separation (Fig. 1D), running South at radii $< 100$ au (PA$\sim 180°$, labeled S1$_{\rm inner}$) and South-East at larger radii (PA$\sim 135°$, labeled S1$_{\rm outer}$). Two broader shadows extend in the North-East (S3) and South-West directions (S4). A filamentary scattered-light structure F$_{\rm scat}$ extends from the inner-most arc (A3) towards the stars (Fig. 1D).

The outer rings R1 and R2 are closely aligned with respect to each other, but strongly misaligned with the orbital plane of the stars, as previously suggested based on disk gas kinematics (*15*). Several physical mechanisms could have produced this misalignment, including turbulent disk fragmentation (*17*), perturbation by other stars in a stellar cluster (*18*), the capture of disk material during a stellar fly-by (*19*), or the infall of material with a different angular momentum vector from the gas that formed the stars (*20, 21*). The innermost ring R3 is strongly misaligned with both the outer disk and the orbits, due to a dynamical interaction with the inner multiple system.

We built a 3-dimensional model, aiming to reproduce both the on-sky projected shape of the dust rings and the shadows seen in scattered light. Based on hydrodynamics simulations (e.g. (*4, 22*)) and the detection of lower-density dust between R2 and R3 in our sub-millimeter image, we modeled this region as a warped dust filament that extends smoothly from ring R2 to a break radius, where the warp is truncated. Our models show (*16*) that material at this inner truncation orbit appears in the scattered light image as the ellipse formed by arcs A3 and A4.

The warped part of the disk facing away from Earth is located South-East of the stars and fully illuminated by them, appearing as arc A3 (Figs. 1C and 2C). The opposite side of the warped disk, located North-West of the stars, is facing towards Earth, so only the outer surface is visible; this is not illuminated by the stars, resulting in the fainter scattered light arc A4. Absorption due to dust in the warped disk reduces the illumination on the North-Western side and causes the broad shadows S3 and S4 at PA$\sim 240°$ and $\sim 20°$, corresponding to the directions with the highest radial column density in the warped part of the disk. The surface of the warped disk also acts as screen for shadows cast by the geometrically thin misaligned ring R3, resulting in the sharply-defined shadow S1. The curvature in S1 can then be understood as a projection effect, where $S1_{inner}$ is the shadow cast on the warped surface inside of R2, while $S1_{outer}$ is the shadow on the non-warped outer disk (Fig. 2A). To reproduce the on-sky projected shape of R3, its off-center position with respect to the stars and the shape of shadows S1 and S2, we adopt a non-zero eccentricity ($e = 0.3 \pm 0.1$ for ring R3), with the stars located at one of the focal points of the ellipse. The Eastern side of ring R3 is tilted away from us, which is consistent with emission from warm ($\sim 70\,\mathrm{K}$) molecular gas that we detect at the inner surface of the ring (Fig. S1, (*16*)). The 3-dimensional orientation of the orbits and dust rings in our model is illustrated in Figs. 2 & 3 and parameterized in Tables S5 & S6.

Observational signatures of broken protoplanetary disks have been predicted in both sub-millimeter thermal emission and near-infrared scattered light (*5*). That work considered a circumbinary disk misaligned by $60°$ with the binary orbit, similar to the misalignment angles observed for GW Orionis ($51.1 \pm 1.1°$ for the A-B orbit and $38.5 \pm 0.8°$ for the (AB)-C orbit). There are similarities between our observations and the predicted synthetic images (*5*), including a misaligned and eccentric ring in sub-millimeter emission and an azimuthal asymmetry in scattered light with sharply-defined shadows. The model eccentricity of ring R3 matches the prediction that the dynamical perturbation by the stars should induce oscillations in the orbital inclination and eccentricity of broken rings (*4, 22, 23*). We compare the radius of R3 (43 au) to analytic estimates of the disk tearing radius, defined as the point in a circumbinary disk where the external torque exerted by a misaligned binary exceeds the internal torque due to pressure forces (*4*). We find that the predicted tearing radius in consistent with the size of R3 for disk viscosity values $\alpha < 0.05$, suggesting that this disk region is susceptible to disk tearing (*16*).

We use the observational constraints on the orbital parameters of the GW Orionis system as input for simulations using smoothed particle hydrodynamic (SPH) and radiative transfer. We parameterize the initial disk with the observed total dust mass (*24*) and adopt the measured stellar orbits and outer disk orientation (*16*). After a few thousand years, the gravitational torque from the misaligned triple star system breaks the disk apart into several distinct planes. An eccentric ring forms with radius $\sim 40$ au which precesses around the inner multiple system with a precession period of $\sim 8,000$ years. Fig. 4 shows a snapshot of this simulation. The size of the simulated ring, its eccentricity, asymmetric azimuthal density profile (with highest density near the farthest part in the ring), and misalignment with the outer disk match the characteristics of ring R3 observed at sub-millimeter wavelengths. This suggests that ring R3 in the GW Orionis system formed by disk tearing. The SPH simulation also forms a low-density warped disk (Fig. 4C), whose properties and spatial orientation broadly resemble the disk warp in our scattered light model (Fig. 2).

The origin of the gap between the two outer-most dust rings seen in millimeter emission (between R1 and R2) remains unclear. The gap might be primarily due to depletion in large dust grains, as mm-sized dust grains might accumulate at the strong density gradient near the outer edge of the disk warp (*25*). Alternatively, the dust gap might coincide with a lower gas density, which could be due to undetected planets within the gap, or disk tearing effects occurring further out in the disk that are not reproduced by our SPH simulation. Some hydrodynamic simulations of misaligned multiple stars find that disk tearing can result in a set of multiple nested rings (e.g. (*4*)) or dust pile-up in warped disk regions resulting from differences in precession between the gas and dust components (*26*), although we estimate that the gas drag forces excerted on the dust particles traced by our millimeter observations are likely too low for the latter mechanism to operate (*16*).

Our results show that disk tearing occurs in young multiple star systems and that it is a viable mechanism to produce warped disks and misaligned disk rings that can precess around the inner binary. By transporting material out of the disk plane, the disk tearing effect could provide a mechanism for forming planets on oblique or retrograde orbits (orbiting in the opposite direction to the orbital axis and/or rotation axes of the stars). About 40% of short-period exoplanets ($\lesssim 40$ days orbital period) are on oblique or retrograde orbits (*27*). The most commonly invoked explanations are planet-planet scattering and tidal interactions from wider-orbiting planets (*28*). Few observations are available for long-period planets and circumbinary planets, with all cases indicating close alignment between the stellar spin and planet orbit plane (*30*) (the most inclined circumbinary planet known is Kepler-413b, with obliquity of $2.5°$ (*29*)). We find that disk tearing can induce large misalignments in a disk, which emerge sufficiently quickly to influence the planet formation process. The broken ring R3 contains $\sim 30$ Earth masses in dust (Table S6), which could suffice for planet formation to occur. Long-period planets on highly oblique orbits could form from rings around misaligned multiple systems.

**Acknowledgments:** This work is based in parts on observations made with European Southern Observatory (ESO) telescopes at the La Silla Paranal Observatory. We thank the ESO Paranal Staff for support for conducting the observations. Atacama Large Millimeter Array (ALMA) is a partnership of ESO (representing its member states), National Science Foundation (NSF, USA) and NINS (Japan), together with NRC (Canada), MOST and ASIAA (Taiwan), and KASI (Republic of Korea), in cooperation with the Republic of Chile. The Joint ALMA Observatory is operated by ESO, AUI/NRAO and NAOJ. The Joint ALMA Observatory is operated by ESO, AUI/NRAO and NAOJ. Based in part upon observations obtained with the Georgia State University (GSU) Center for High Angular Resolution Astronomy (CHARA) Array at Mount Wilson Observatory. The CHARA Array is supported by the NSF under Grant No. AST-1636624 and AST-1715788. Institutional support has been provided from the GSU College of Arts and Sciences and the GSU Office of the Vice President for Research and Economic Development. Based in part on observations obtained at the Gemini Observatory, which is operated by the Association of Universities for Research in Astronomy, Inc., under a cooperative agreement with the NSF on behalf of the Gemini partnership: the NSF (United States), the National Research Council (Canada), CONICYT (Chile), Ministerio de Ciencia, Tecnologıa e Innovacion Productiva (Argentina), and Ministerio da Ciencia, Tecnologia e Inovacao (Brazil). Some figures were produced using the SPH visualization tool SPLASH. The SPH and radiative transfer calculations discussed in this paper were performed on the University of Exeter Supercomputer, Isca.

**Funding:** S.K., A.K., and C.L.D. acknowledge support from the European Research Council (ERC) under the European Commission's (EC) Horizon 2020 program (Grant Agreement Number 639889). A.K.Y. acknowledges funding from the ERC under the EC's Horizon 2020 program (Grant Agreement Number 681601). M.R.B. acknowledges support from the ERC un-der the EC's Seventh Framework program (Grant Agreement Number 339248). A.L. thanks the Science Technology and Facilities Council (STFC) for a studentship that supported this work (project reference 1918673). J.D.M. and E.A.R. acknowledge funding from NSF grant NSF-AST1506540, NSF-AST 1830728, and NASA grant NNX16AD43G. J.K. acknowledges support from the research council of the KU Leuven under grant number C14/17/082. J.B. acknowledges support by NASA through the NASA Hubble Fellowship grant #HST-HF2-51427.001-A awarded by the Space Telescope Science Institute, which is operated by the Association of Universities for Research in Astronomy, Incorporated, under NASA contract NAS5-26555.

**Author contributions:** S.K. conceived the project, initiated the ALMA, Very Large Telescope (VLT), Very Large Telescope Interferometer (VLTI), and CHARA observing programs, modeled the infrared and sub-millimeter data, fitted the triple star orbit, and wrote the initial manuscript. A.K. implemented the scattered light model and processed the ALMA data. A.K.Y. and M.R.B. performed the radiative transfer and SPH simulations. J.D.M. initiated the Gemini Planet Imager (GPI) observing program. H.A. and J.K. processed the VLT scattered-light imaging data sets. A.S.E.L. and E.A.R. processed the GPI data set. S.K., J.D.M., N.A., C.L.D., J.E., T.G., A.L., C.L., J.-B.L.B. G.H.S., and B.R.S. built and commissioned the Michigan InfraRed Combiner-eXeter (MIRC-X) instrument for CHARA. S.K., J.D.M., T.J.H., A.N.A., F.C.A., S.M.A., J.B., N.C., C.E., L.H., S.H., D.W., and Z.Z. contributed to the GPI survey. All co-authors provided input on the manuscript.

**Competing interests:** There are no conflicts of interest.

**Data and materials availability:** The VLT and VLTI data are archived in the ESO Science Archive (http://archive.eso.org) under program IDs 082.C-0893(A), 384.D-0482(A,B), 084.C-0848(A-G), 086.C-0684(A,B), 088.C-0868(A), 090.C-0070(A), 094.C-0721(A), 098.C-

0910(A), 100.C-0686(A-C), and 102.C-0778(A). The ALMA data is archived in the ALMA Science Archive (http://almascience.nrao.edu/aq/) under project codes ADS/JAO.ALMA #2012.1.00496.S and #2018.1.00813.S. The reduced data cubes, model files, and simulation codes are available from https://sourceforge.net/projects/aba4633/. GPI data are accessible from the Gemini Observatory archive (https://archive.gemini.edu) under program ID GS-2017B-LP-12.

**Supplementary materials:**   Materials and Methods

Supplementary Text

Figs. S1 to S13

Tables S1 to S6

References *(31-85)*

**Fig. 1. Imaging of the disk components around GW Orionis.** (A) and (B) shows the 1.3 mm thermal dust continuum emission on different spatial scales, measured in the spectral flux density unit milli-Jansky (mJy). The main components seen in the images are labelled, including three rings (R1,R2,R3), an asymmetry in the ring R3 (R3$_{\text{asym}}$), and dust emission close to the stars (D$_{\text{AB}}$, D$_{\text{C}}$). (C) and (D) shows near-infrared (C) and visible-wavelength (D) scattered light, where the images have been multiplied with $r^2$ to emphasise structures in the outer disk, where $r$ is the distance from the stars in the image. Four arc structures (A1,A2,A3,A4) and a filamentary structure (F$_{\text{scat}}$) are labeled. There are four radial shadows (S1,S2,S3,S4); S1 changes orientation, with a different position angle within and outside 100 au (S1$_{\text{inner}}$ and S1$_{\text{outer}}$, respectively). In panels B-D the orbits and positions of the stars at the time of observation are indicated by blue (GW Ori A), orange (GW Ori B), and white (GW Ori C) curves and symbols. We indicate the angular resolution (beam) achieved by the observation. In all panels, North is up and East is left as indicated in panel B.

**Fig. 2. Scattered light model (*16*) used to determine the eccentricity and 3-dimensional orientation of ring R3 and the geometry of the disk warp.** (A) Diagram of the 3-dimensional orientation of the disk components in the model. (B) Orientation of the orbits in the hierarchical triple star system, with the stellar positions at the time of our imaging measurements indicated (same as in Fig. 1). The coordinate system is centered on the center-of-mass of the system. The white grid indicates the observed plane of the sky (right ascension RA and declination Dec, while the z-axis points towards the observer, represented by the eye symbol in panel A. (C) Synthetic image computed (*16*) from the scattered light model, convolved to match the resolution of the observation. The position of the scattered light features from Figs. 1C-D are indicated.

**Fig. 3. Interactive 3-dimension model.** The model can be rotated in Adobe Reader to display the disk geometry. The x and y axes correspond the direction of Dec (North) and RA (East), while z axes points towards Earth. Zooming in shows the triple star orbits (GW Ori A: blue, GW Ori B: orange, and GW Ori C: black).

**Fig. 4. Smoothed particle hydrodynamic simulation.** The computation is based on the measured GW Orionis orbits and system parameters, evolved for 9500 years. (A) gas density projected on the plane of the sky, with North up and East left. The third axis (positive z) is facing out of the page. (B) and (C) integrated gas density projected in the z-Dec plane and RA-z plane. (D) density cut along the RA-z plane.

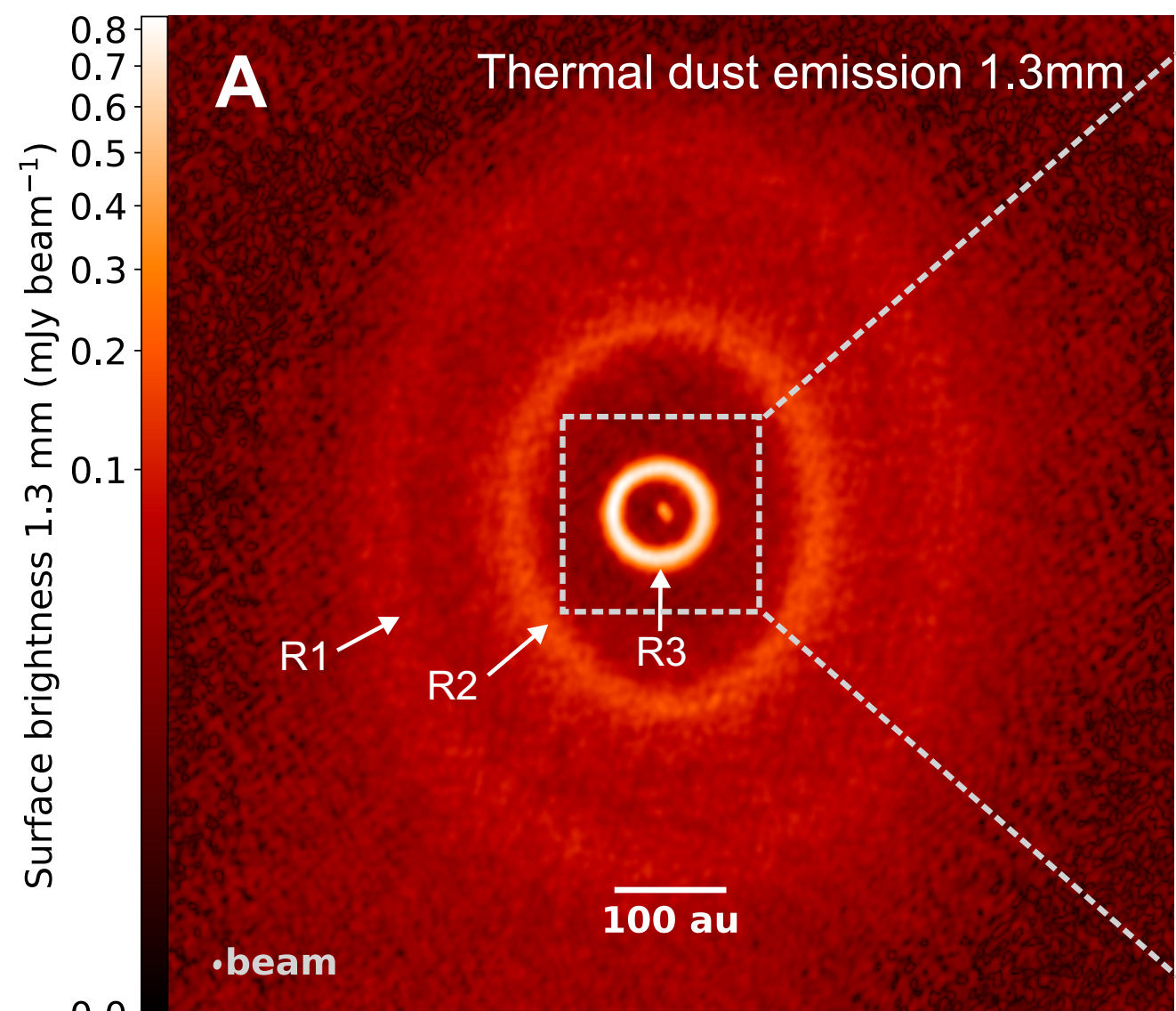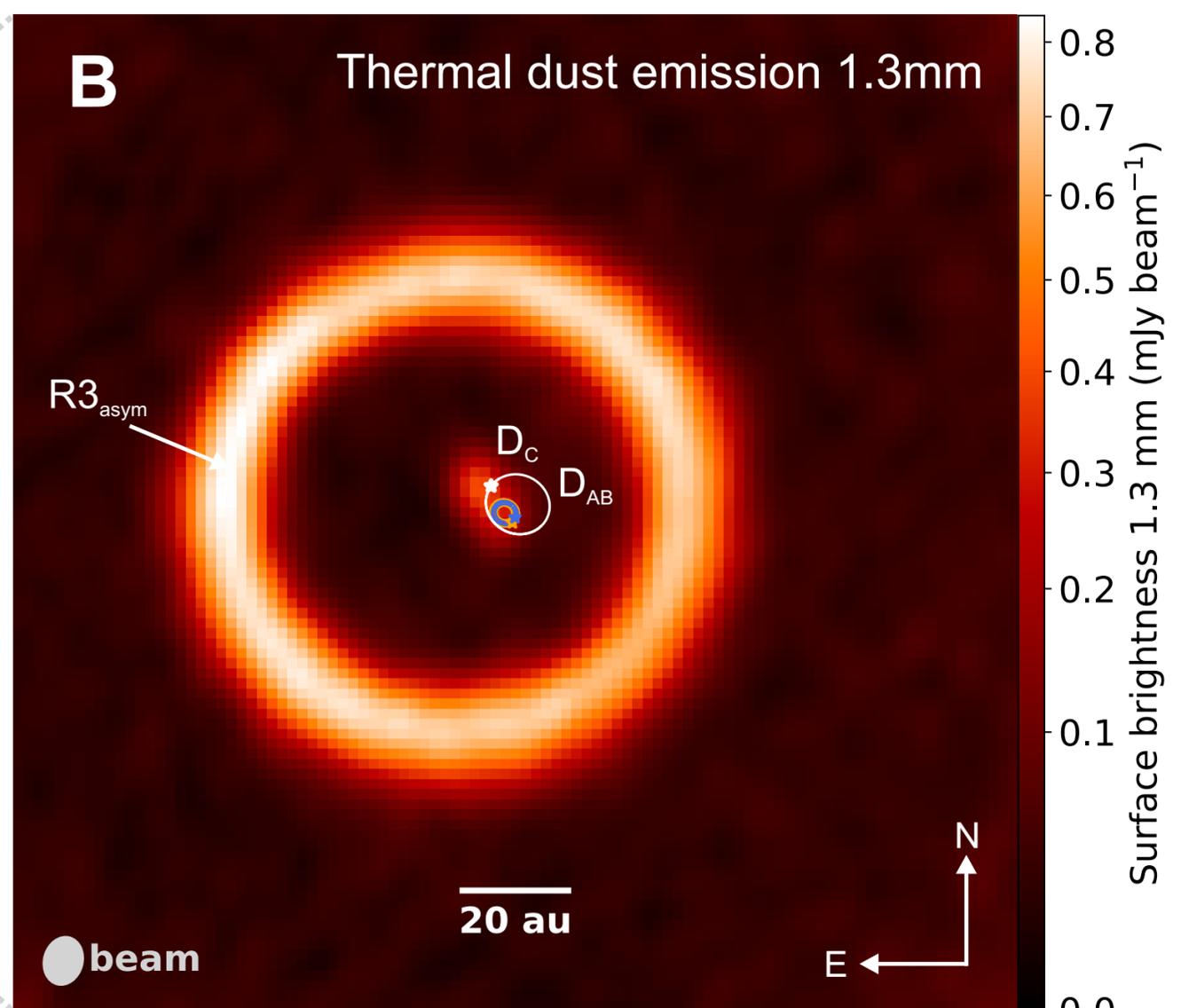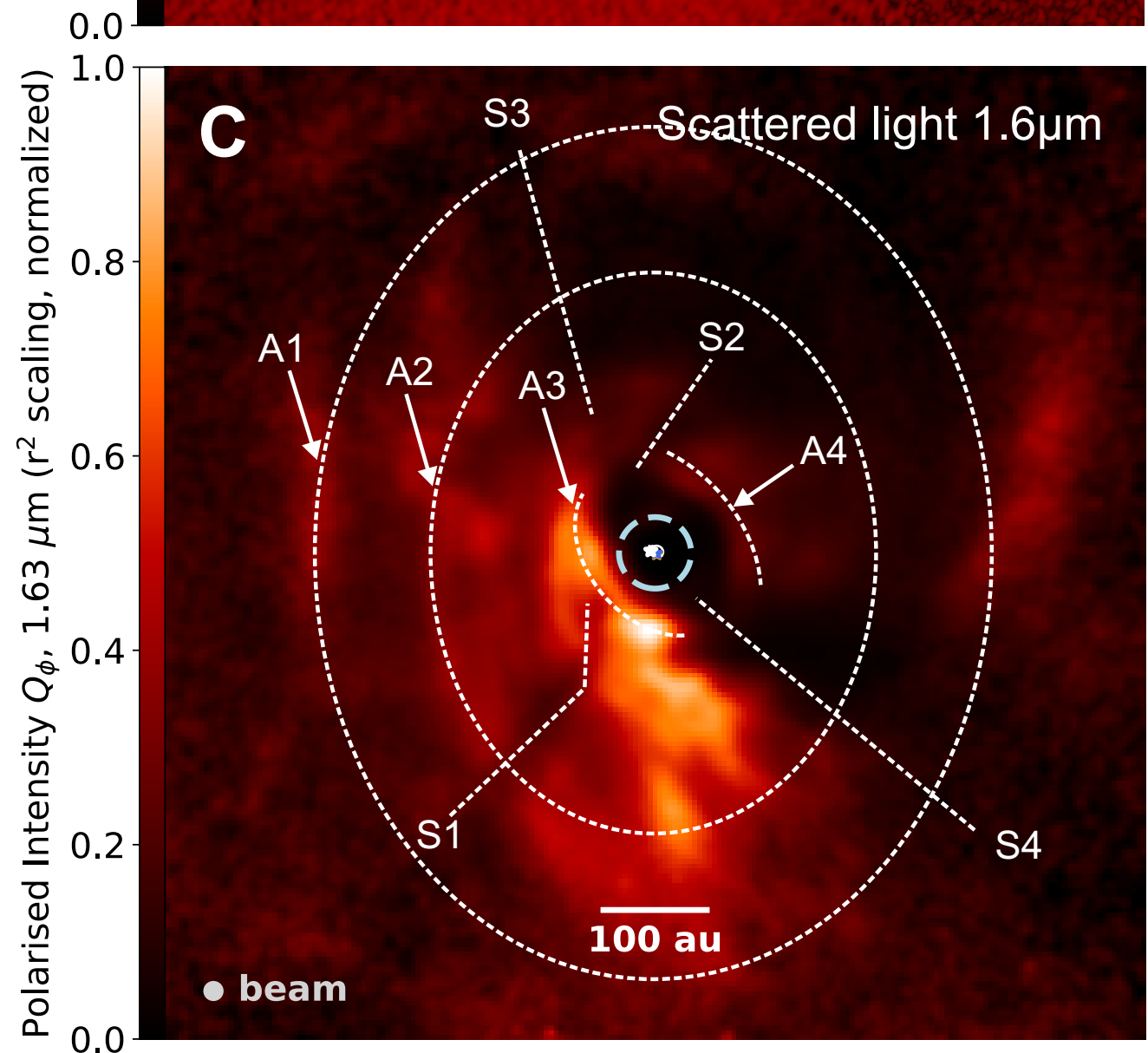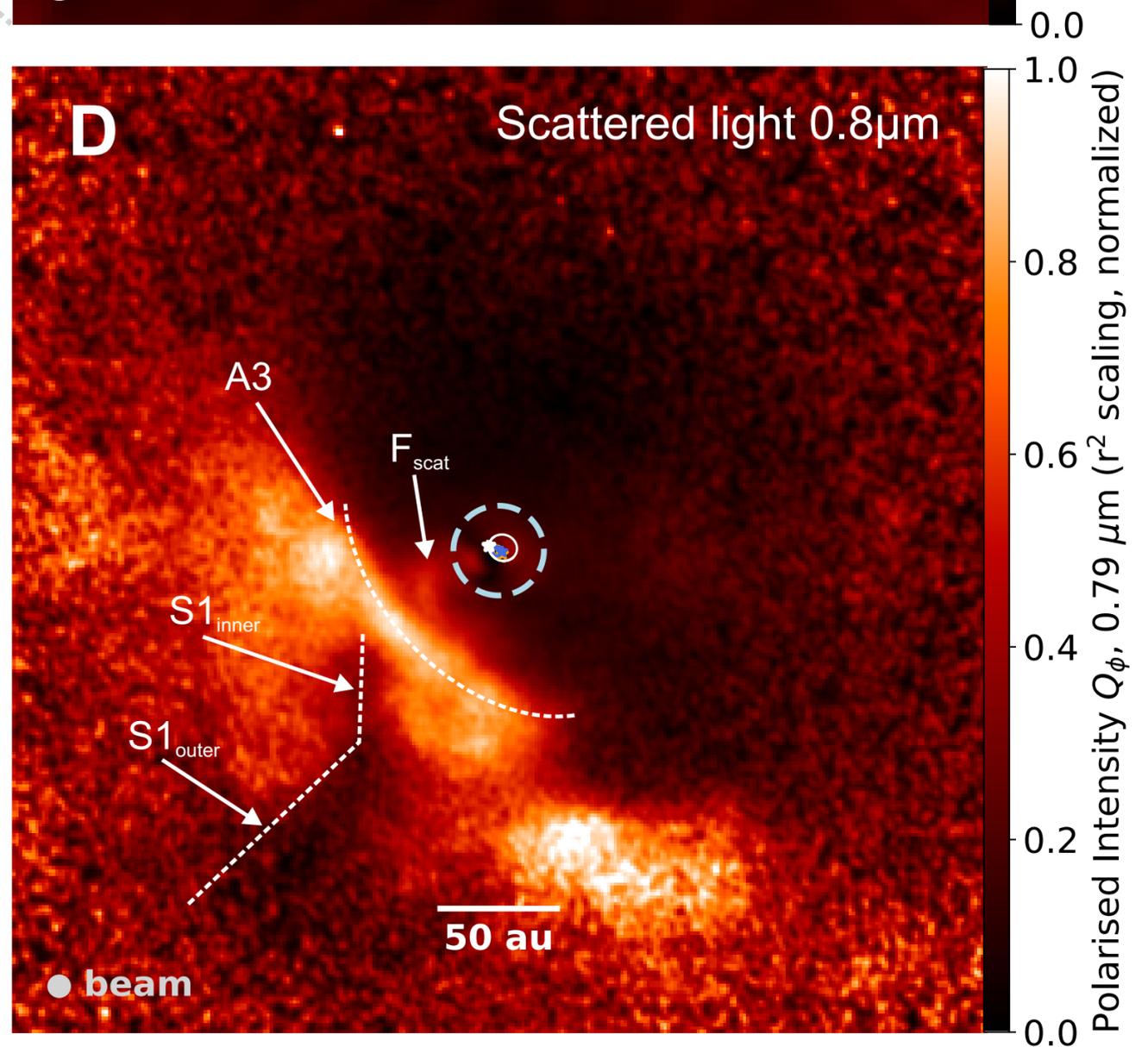

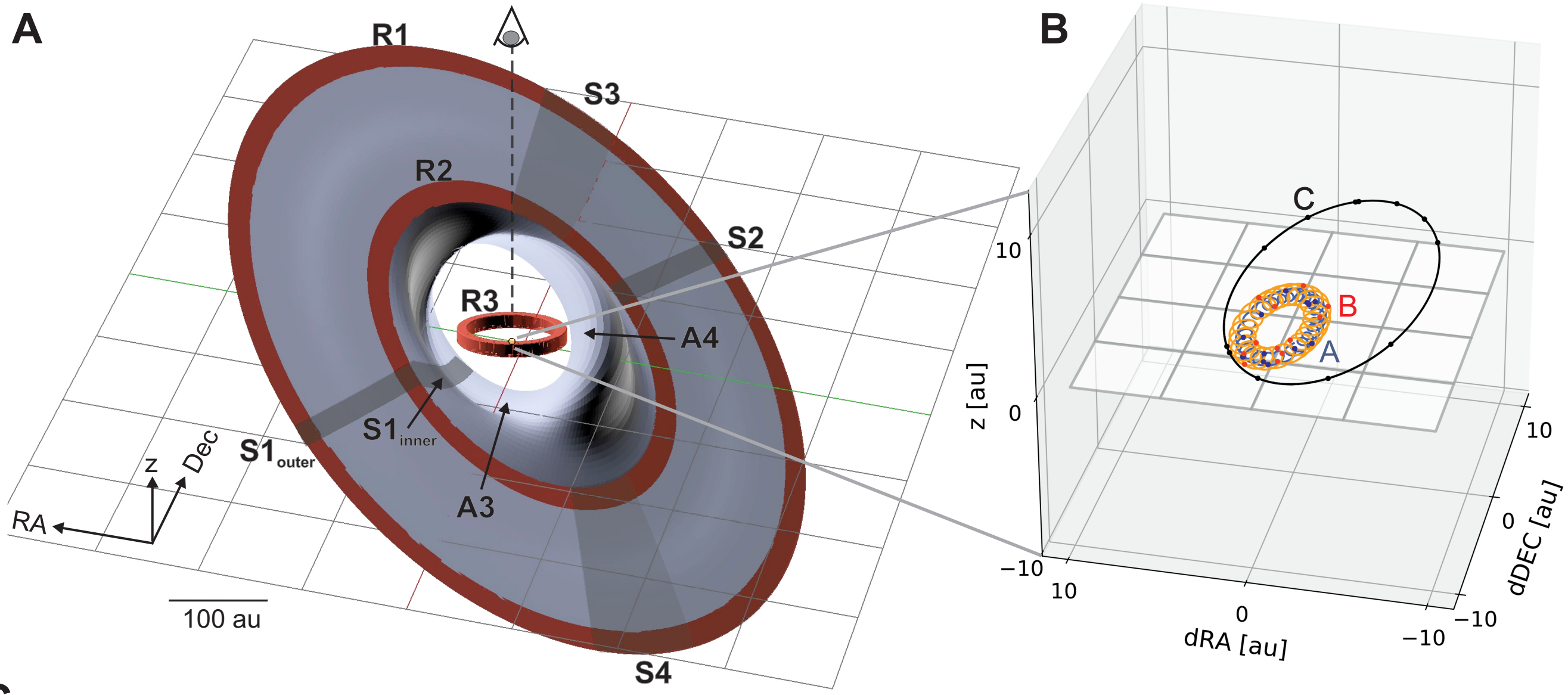

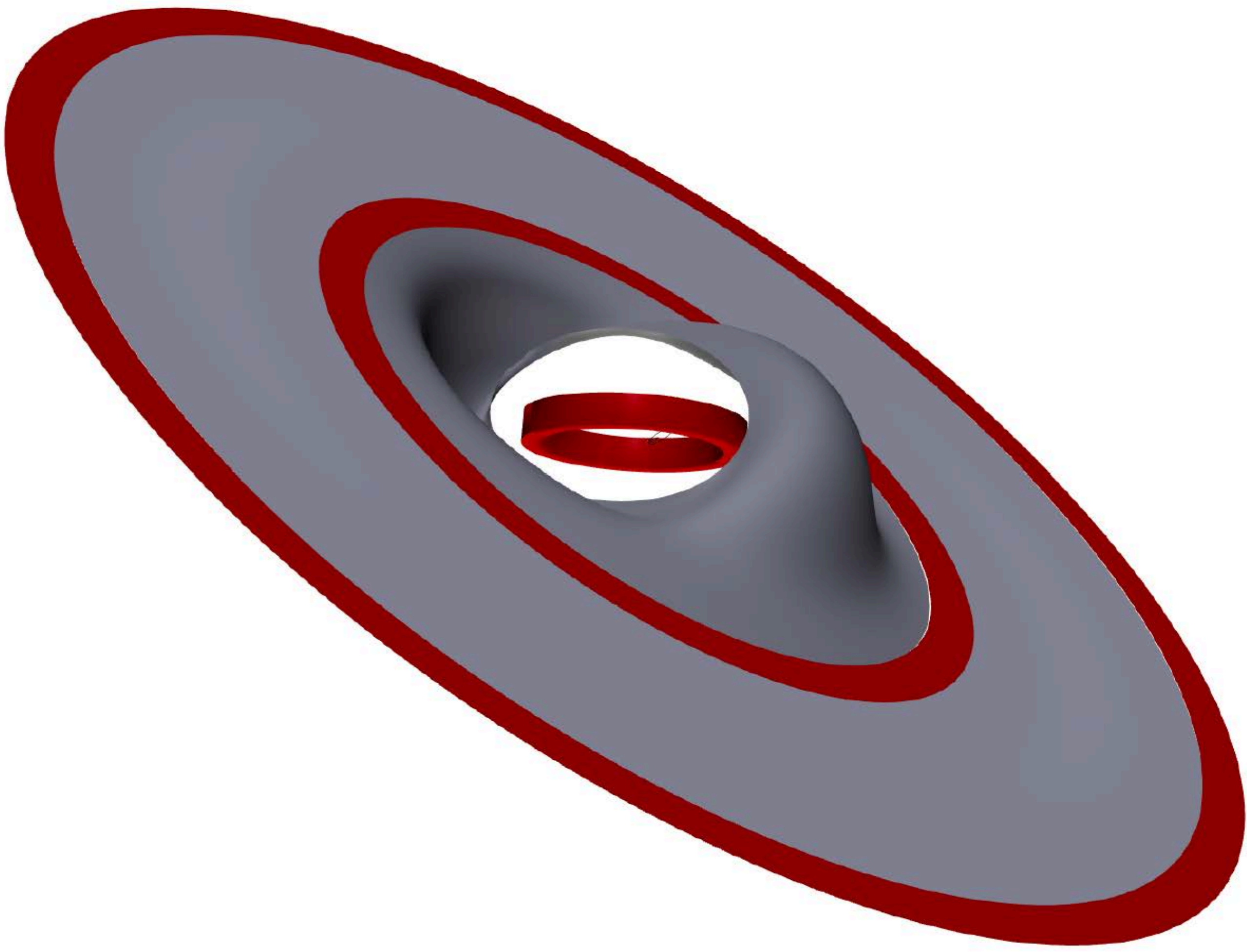

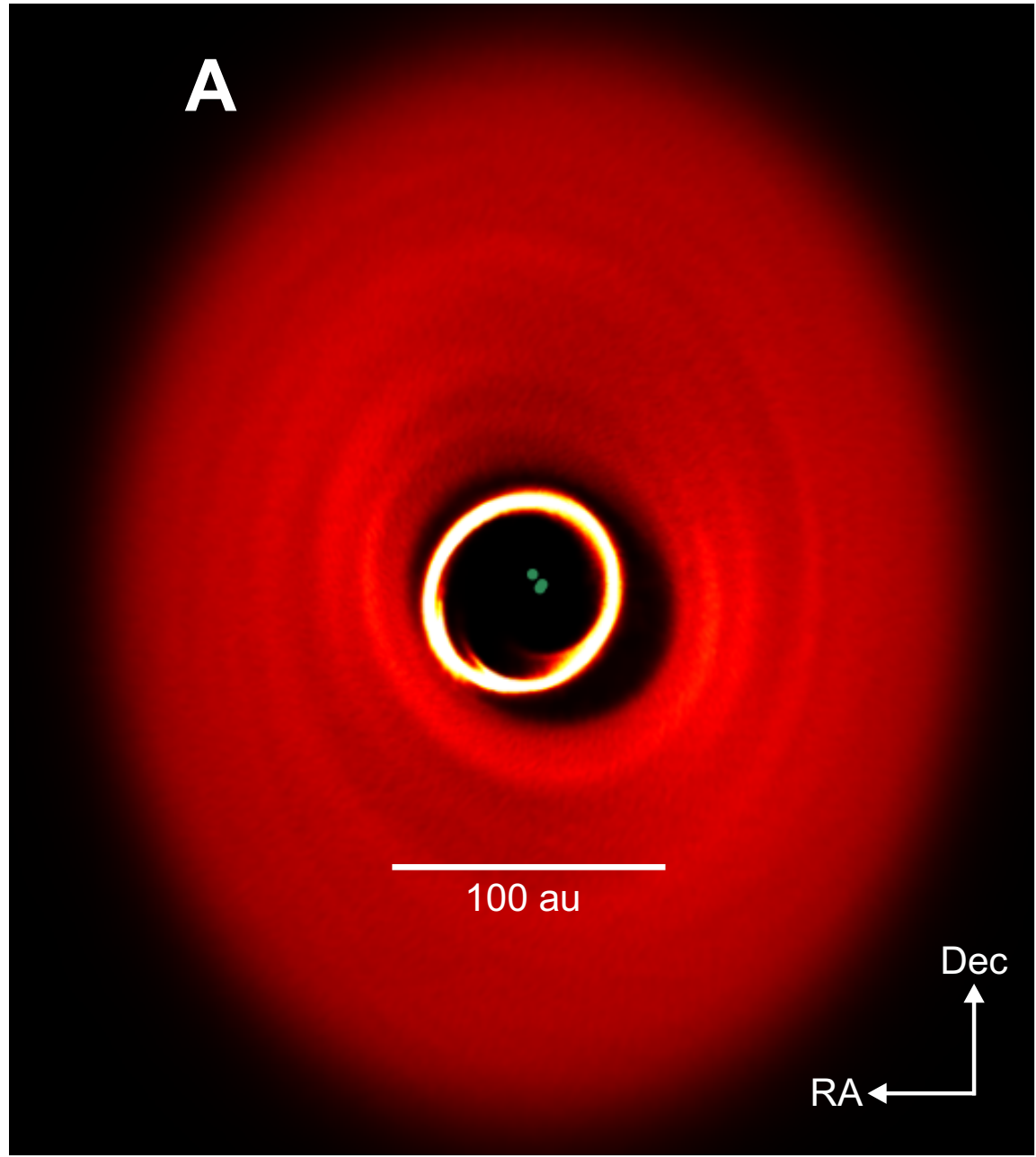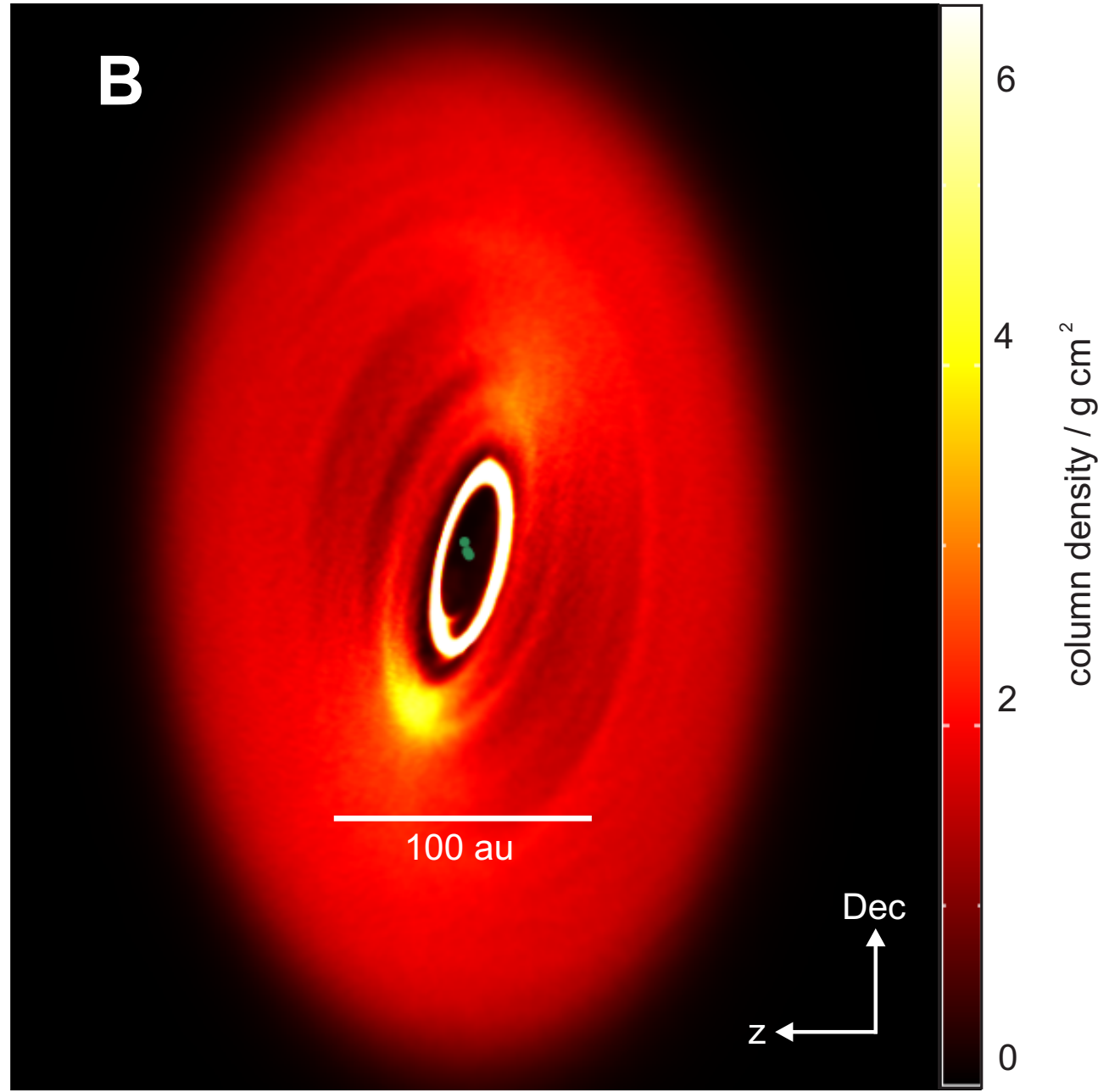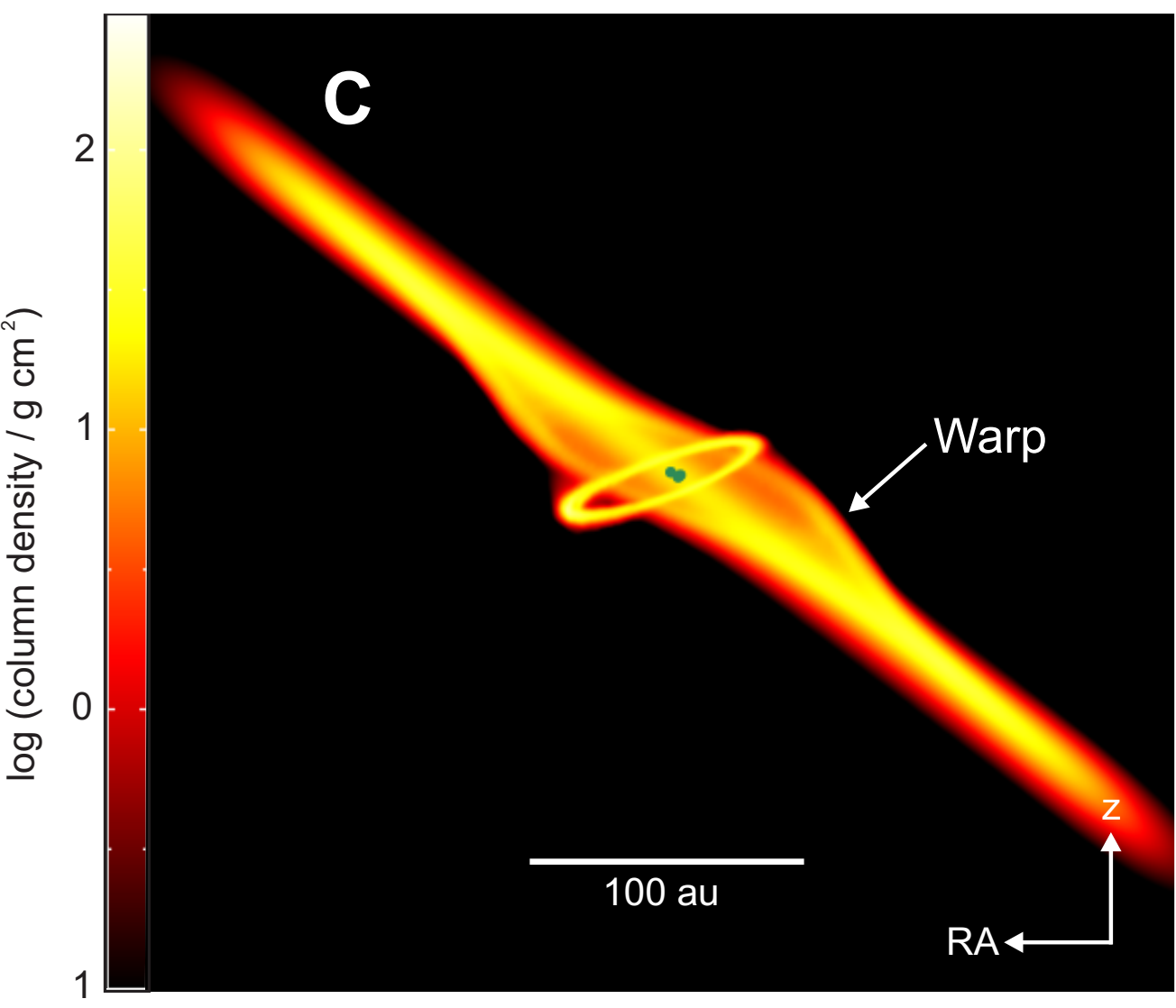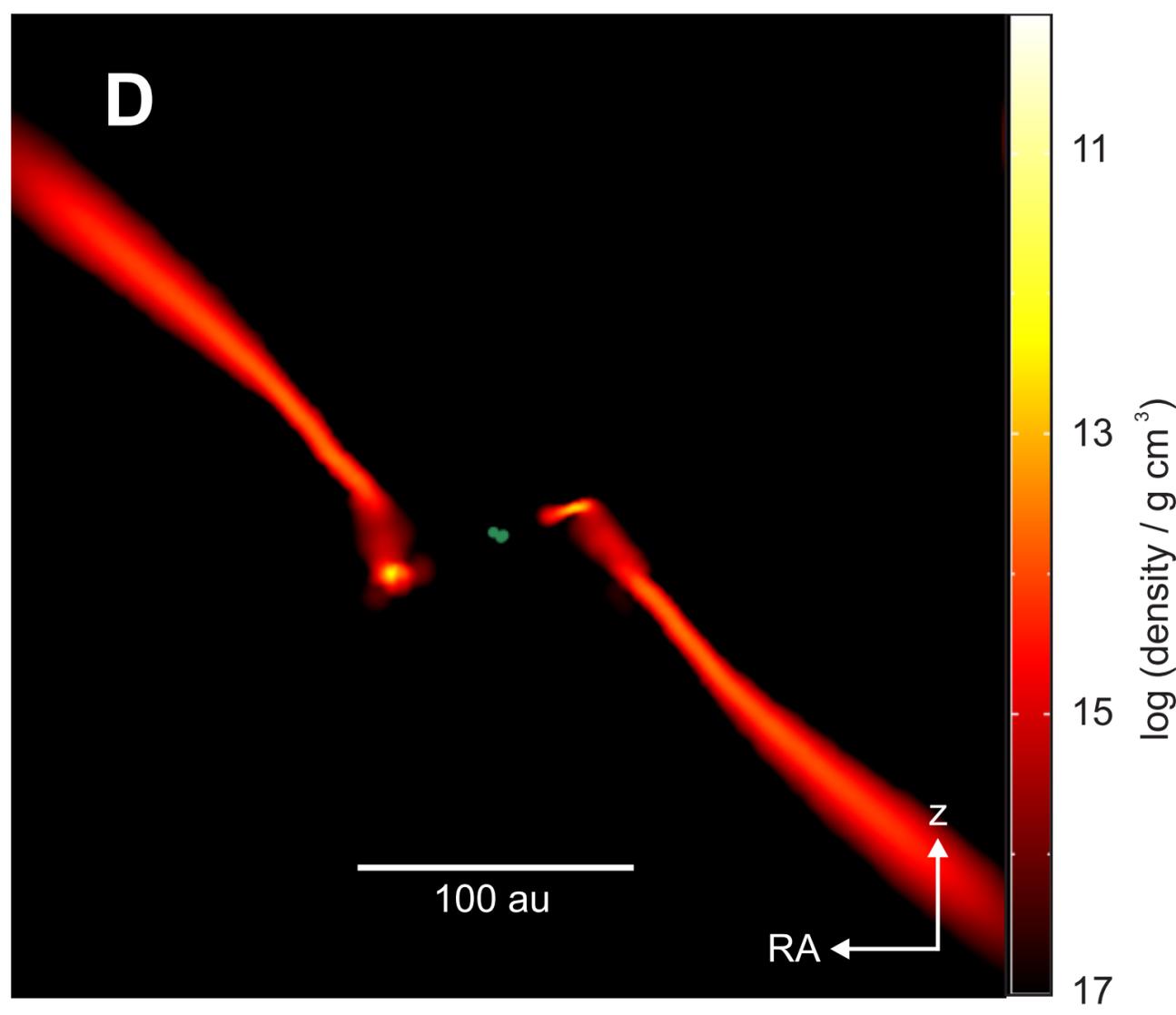

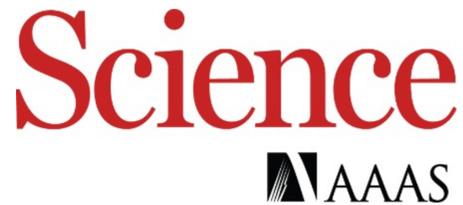

Supplementary Materials for

A triple star system with a misaligned and warped circumstellar disk shaped by disk tearing

Stefan Kraus, Alexander Kreplin, Alison K. Young, Matthew R. Bate,
John D. Monnier, Tim J. Harries, Henning Avenhaus, Jacques Kluska,
Anna S. E. Laws, Evan A. Rich, Matthew Willson, Alicia N. Aarnio,
Fred C. Adams, Sean M. Andrews, Narsireddy Anugu, Jaehan Bae,
Theo ten Brummelaar, Nuria Calvet, Michel Curé, Claire L. Davies,
Jacob Ennis, Catherine Espaillat, Tyler Gardner, Lee Hartmann,
Sasha Hinkley, Aaron Labdon, Cyprien Lanthermann, Jean-Baptiste LeBouquin,
Gail H. Schaefer, Benjamin R. Setterholm, David Wilner, Zhaohuan Zhu

Correspondence to: s.kraus@exeter.ac.uk

**This PDF file includes:**

Materials and Methods
Supplementary Text
Figs. S1 to S13
Tables S1 to S6

# S1 Materials and Methods

## S1.1 Observations

The data used for this study is summarized in Table S1. In the following sections, we provide details on the instrument setups that were used and the data reduction process. The infrared interferometric observations on GW Ori (right ascension 05h 29m 08.3925s; declination +11:52:12.654, J2000 equinox) used the *Astronomical Multi-BEam combineR* (AMBER), Gravity, and *Michigan InfraRed Combiner-eXeter* (MIRC-X) instruments and were interleaved with observations on calibrator stars of known uniform disk diameter $d_{\rm UD}$. The properties of these calibrators are listed in Table S2.

### S1.1.1 VLTI/AMBER near-infrared interferometry

We observed GW Ori with the AMBER instrument (*31*) and ESO's Very Large Telescope Interferometer (VLTI). The observations were conducted at seven epochs between 2008 December 15 and 2015 January 1 (Table S1) and used three 8.2 m unit telescopes, resulting in projected baseline lengths between 24.4 and 129.9 m. The data covers wavelengths between 1.4 and 2.5 μm (H- and K-band) with resolving power $\lambda/\Delta\lambda = 35$, where $\lambda$ is the observing wavelength and $\Delta\lambda$ is the width of each spectral channel. For the AMBER observations the achieved angular resolution was $\lambda/2B_{\rm max} \approx 1.4$ milliarcseconds (mas) with a field of view of $\sim 60$ mas, where $B_{\rm max}$ is the projected length of the longest baseline used for the observations.

We extract wavelength-dependent visibilities and closure phases from the data using the AMDLIB software, release 3 (*32, 33*). To minimize the effect of residual telescope jitter, we follow the standard AMBER data reduction procedure and select the best 10% of interferograms with the highest signal-to-noise ratio (*34*). We reject scans where the optical-path-delay exceeded 4 μm to improve the visibility calibration. The wavelength calibration was done using the telluric absorption bands between the J, H, and K-bands [(*35*), their appendix A].

### S1.1.2  VLTI/Gravity near-infrared interferometry

On three nights in 2017 and 2018 we acquired observations with the VLTI/Gravity instrument (*36*). The observations combined the light from the four VLTI 1.8 m auxiliary telescopes and covered the K-band (1.9 to 2.5 μm) with $\lambda/\Delta\lambda = 500$. For the 2017 observations, the telescopes were in the compact configuration (A0-B2-D0-C1), corresponding to projected baselines in the range 8.4 to 31.4 m. With an angular resolution of $\lambda/2B_{\mathrm{max}} = 7.0$ mas, the 2017 Gravity observations did not resolve the inner binary (A-B) pair and the data set was therefore not included in our further analysis.

The 2018 observation was conducted with the extended configuration (A0-G1-J2-J3). This resulted in projected baseline lengths between 57.5 and 129.4 m and resolution $\lambda/2B_{\mathrm{max}} = 1.7$ mas, providing astrometric constraints for all stars. The field of view for the Gravity observations was $\sim 100$ mas.

Wavelength-differential visibilities and phases were extracted using the Gravity pipeline (Release 1.2.4, (*37*)).

### S1.1.3  CHARA/MIRC-X near-infrared interferometry

In 2019 we acquired another epoch of near-infrared interferometry (1.45 to 1.8 μm; H-band) using the MIRC-X instrument (*38, 39*). Based on the *Michigan InfraRed Combiner* (*40, 41*), this instrument combines light from all six telescopes of the *Center for High-Angular Resolution Astronomy* (CHARA) array simultaneously, resulting in visibility measurements on 15 baselines and 20 closure phase measurements in a single pointing. Our data were obtained on projected baseline lengths between 32.0 and 330.7 m, which improved the angular resolution and astrometric accuracy by a factor 2.5 compared to our earlier VLTI observations. The data were recorded using the PRISM50 spectral mode, which covers the 1.5-1.72 μm range (H-band) with resolving power $\lambda/\Delta\lambda = 50$. The observations achieved an angular resolution of $\lambda/2B_{\mathrm{max}} = 0.5\,\mathrm{mas}$ with a field of view of $\sim 300\,\mathrm{mas}$.

For the data reduction, we used the MIRC-X data reduction pipeline (*42*).

**S1.1.4  ALMA sub-millimeter interferometry**

Our ALMA band 6 data were recorded on 2019 July 5 as part of ALMA Cycle 6 program 2018.1.00813.S. The array included 44 antennas arranged in configuration C43-9/10, covering baseline lengths between the $B_{\mathrm{min}} = 241.0\,\mathrm{m}$ and $B_{\mathrm{max}} = 10,240.9\,\mathrm{mm}$, where $B_{\mathrm{min}}$ and $B_{\mathrm{max}}$ are the projected length of the shortest and longest baseline used for the observation.

The observations cycled between GW Ori and the phase calibrator QSO J0530+13. In total, 46.4 min data were recorded on-source for GW Ori. The bright quasar QSO B0507+179 was used as bandpass and flux calibrator, while ICRF J053942.3+143345 was observed as check source. Our setup included four spectral basebands centered on 230.55, 232.5, 217.5, and 219.465 GHz. Each spectral band covers a bandwidth of 1875 MHz with a spectral resolution of 1.129 MHz. Our correlator setup covered the Carbon Monoxide $^{12}\mathrm{CO}\,(J = 2 \rightarrow 1)$ line transition (rest frequency 230.538 GHz), where $J$ indicates the rotational quantum number.

The standard flagging and calibration was done using the *Common Astronomy Software Applications* (CASA) package, version 5.6.0 (*43*). Self-calibration was performed with the shortest phase and amplitude cycles of 10 and 60 seconds, respectively. For the image reconstruction process we used only line-free spectral channels and employed the Clean algorithm with a Briggs weight of 0.5, which resulted in a beam size of $24 \times 19\,\mathrm{mas}$ (along PA $-13°$) and a root-mean-square noise of $60\,\mu\mathrm{Jy\,beam}^{-1}$.

The extended array configuration C43-9/10 means our observations lack sensitivity to extended emission. The maximum recoverable scale is $0.6\lambda/B_{\mathrm{min}} \approx 0.67''$ (*44*). To retrieve flux on larger angular scales we combine our ALMA data with short-baseline data from program 2012.1.00496.S that spans baseline lengths between 23 and 558 m. We performed self-calibration on the two continuum data sets separately, before combining them into a single measurement set that was then used for image reconstruction. The flux density measured in our image are 340 mJy and 165 mJy for a 2″ and 1″ circular aperture centered on the position of the star. These values are similar to $320 \pm 64$ mJy (*45*) and $202 \pm 20$ mJy (*15*) measured with short-baseline SMA and ALMA data for the same aperture sizes, indicating that our image captures extended emission and does not suffer from spatial filtering.

To improve the signal-to-noise for imaging in the $^{12}$CO $(J = 2 \to 1)$ emission line we rebinned the CO channel maps to 40 mas resolution and used to CASA task IMMOMENTS to retrieve the CO surface brightness (zeroth-moment) map and intensity-weighted velocity (first-moment) map shown in Fig. S1.

### S1.1.5 GPI near-infrared polarimetric imaging

Imaging polarimetric observations were conducted with the Gemini Planet Imager (GPI; (*46*)) on 2018 January 4 using both the "J-coron-pol" (J-band; $\lambda_f = 1235$ nm; $\Delta\lambda_f$=230 nm) and "H-coron-pol" (H-band; $\lambda_f = 1650$ nm; $\Delta\lambda_f$=300 nm) instrument setup, where $\lambda_f$ and $\Delta\lambda_f$ are the central wavelength and width of the employed broadband filter. We utilized the standard GPI coronagraphic masks that have diameters of 0.″184 (for J-band) and 0.″246 (H-band).

The data was reduced using the GPI Data Reduction Pipeline (DRP), version 1.5 (*47*). We follow established procedure (*48, 49*) and project the traditional Stokes parameter $Q$ and $U$ onto an azimuthal set of Stokes parameters, $Q_\phi$ and $U_\phi$. This representation of the Stokes vector yields positive $Q_\phi$ values when the field polarization angle is perpendicular to the vector connecting the pixel and the star's location, while radial vectors are negative. We do not attempt to minimize $U_\phi$ as it could be non-zero for disks seen at high inclination (*49*). However, for most cases of single-scattering and moderate inclination angles, $U_\phi$ is expected to remain free of astrophysical signals. Fig. S2 shows the Stokes $Q_\phi$ and $U_\phi$ images. We focus our interpretation on the $Q_\phi$ maps, as they measure the azimuthally polarized flux from the disk.

### S1.1.6 SPHERE visible + near-infrared polarimetric imaging

Our polarimetric imaging with the *Spectro-Polarimetric High-contrast Exoplanet REsearch* (SPHERE) instrument (*50*) utilized both the near-infrared arm (*Infra-Red Dual-band Imager and Spectrograph*, IRDIS, (*51*)) and the visible-light arm (*Zurich IMaging POLarimeter*, ZIMPOL, (*52*)). The IRDIS observations were conducted with the H-band filter (BB_H, $\lambda_f = 1625$ nm; $\Delta\lambda_f$=290 nm) and the smallest available coronagraphic mask (N_ALC_YJ_S; inner working angle 0.08"). Images for nine polarimetric cycles were recorded, each with six images of 16 s integration time. The ZIMPOL observations were conducted with seeing conditions of 0.3-0.4" full-width-at-half-maximum (FWHM) and using the I'-band filter (I_PRIM, $\lambda_f = 789.7$ nm; $\Delta\lambda_f$=152.7 nm). We did not use a coronagraphic mask but required a neutral density filter to avoid saturation (ND_2.0 filter). Six polarimetric cycle sequences were recorded, each recording 4 images of 8 s integration time for each cycle step.

For the data processing of the IRDIS data, we use the IRDAS software package (*53*) that follows the procedures outlined in (*54*) and derives Stokes $Q_\phi$ and $U_\phi$ maps. To extract the polarized intensity images from ZIMPOL data, we applied the procedure outlined in (*55*), deriving again $Q_\phi$ and $U_\phi$. The resulting images are shown in Fig. S3.

### S1.1.7 Archival astrometric and radial velocity data

To further constrain the orbit we incorporate published radial velocity (RV) measurements (*15*). We use the measurements that were corrected for instrumental effects. This data covers the period from 1981 November 11 to 2017 April 20 at 284 epochs.

We also incorporate published visibility and closure phase data (*14*). Their data was obtained with 3 telescopes from the Infrared Optical Telescope Array (IOTA) and covers three epochs, 2003 November 30/December 1, 2004 December 12-20, and 2005 November 22 with projected baseline lengths between 7.9 m and 36.7 m.

## S1.2 Modeling

For the stellar parameters of GW Ori, we adopt the values listed in Table S3 throughout.

### S1.2.1 Triple star astrometry and circumbinary disk R4

For a first, model-independent inspection of our near-infrared interferometric observations we applied aperture synthesis imaging methods to our VLTI and CHARA data. We employed the SQUEEZE (*56, 57*) image reconstruction software and used a $600 \times 600$ pixel grid with a pixel scale of 0.1 mas, and a total variation regularizer with a weight of 2000. Representative images for two epochs are shown in Fig. S4.

For each epoch of our VLTI/AMBER+Gravity observations, we derive the relative astrometry of the GW Ori system using a triple star model, as implemented in the LITPRO software (*58*). This model is similar to the one adopted to interpret the IOTA data (*14*). However, we find that such a 3-point source model provides a poor fit to our VLTI+CHARA data, which provides up to 10-times higher spatial frequencies than the earlier IOTA observations. Specifically, the point-source model predicts much higher visibilities than measured on long baselines. We therefore introduce an extended emission component that contributes $F_{\text{ext}}/F_{\text{tot}} = 16 \pm 2\%$ of the total flux. For the geometry of this extended emission component, we first considered an overresolved halo, as is routinely adopted to reproduce extended flux in T Tauri and Herbig Ae/Be stars (e.g. (*59, 60*)). Such a model is able to reproduce the VLTI data. However, when we apply this model to the 2003-2005 IOTA data or the short-baseline Gravity data, we find that the overresolved flux lowers the visibilities too much on short ($\lesssim 15$ m) baselines, resulting in a very poor fit. This constrains the size of the extended component, as the emission needs to be sufficiently compact to be only marginally resolved on short baselines, but fully resolved on baselines $\gtrsim 30$ m. Therefore, the extended emission component is more complex than a Gaussian or uniform intensity profile and we use the combined IOTA+VLTI data to constrain the geometry of this extended emission. We fitted rings with uniform brightness and smoothed rings, where the radial ring intensity profile is parameterized as a Gaussian with fixed FWHM of 0.5 mas, to avoid unphysical sharp edges in the brightness profile. We find that a smooth ring centered on the A component (in the following referred to as R4) provides the best representation of this extended flux component and results in $\chi^2_{\rm red} = 1.6$, compared to $\chi^2_{\rm red} = 6.3$ for an overresolved halo geometry. This modeling shows the extended near-infrared flux component has a radius of $\sim 5^{+7}_{-2}$ mas, equivalent to $2^{+2.5}_{-0.8}$ au, likely tracing hot dust at the inner edge of a circumbinary disk around the A+B component. Our observing setup was optimized for precision astrometry at high spatial frequencies, so we lack the $(u, v)$-plane coverage in the short-baseline regime needed to characterize this extended near-infrared emission component further and to determine its precise inclination and orientation in space ($u$ and $v$ refer to the sky-projected coordinates of the baseline vector). Our visibility modeling places an upper limit of 2% on the H-band flux contributions for any over-resolved halo or extended scattered light component within an field-of-view of $\sim 0.2''$, including total intensity and polarized light.

When fitting the IOTA+VLTI data, we represent the stars as point sources. For the long-baseline CHARA/MIRC-X data we find that the visibility is systematically lower than predicted by the model with point sources, resulting in $\chi^2_{\rm red} = 5.9$ (where $\chi_{\rm red}$ is the reduced goodness-of-fit indicator, as defined in (*35*)). The model was improved by assuming that components A and B are marginally resolved, with $\chi^2_{\rm red} = 3.6$, adopting uniform disk diameters of $0.6$ mas and $0.57$ mas. This corresponds to 25 and 24 solar diameters, respectively, at the distance of GW Ori, indicating that we resolve some circumstellar dust, possibly arranged in small circumstellar disks. Including resolved emission around the C component does not improve the significance of the fit.

Figs. S5, S6, and S7 show the calibrated visibilities and phases for three representative epochs, overplotted with the best-fitting model. The fits to the Gravity 4-telescope data (2018.099) and MIRC-X 6-telescope data (2019.652) feature slightly larger $\chi^2_{\rm red}$ values than the fits to the IOTA and AMBER 3-telescope data. This is due to closure phase residuals that occur near phase jumps (Fig. S7). Near these jumps, the measured closure phase profiles are very sensitive to small details in the brightness distribution and to minor instrumental effects (such as edge effects between the spectral channels or uncertainties in wavelength calibration).

**S1.2.2 Astrometric+spectroscopic orbit solution**

The existing IOTA+VLTI+CHARA astrometry covers 12 epochs over 15.8 years, or 1.35 full orbital periods of the outer orbit (∼23 periods of the inner orbit). A previously-published astrometric orbit solution (*15*) was based on data from 3 epochs, covering 17% of the orbit of the outer component. Their astrometry at the third epoch was degenerate due to $(u, v)$-coverage limitations, leading to two alternate solutions. Our VLTI+CHARA observations provide astrometry at nine additional epochs, which allows us to resolve these ambiguities and to constrain the 3-dimensional orbits of all stars in the system. Our data have 10-times higher angular resolution and denser $(u, v)$-coverage, which allows us to detect an extended emission component that corresponds to the circumbinary ring R4 (see Sect. S1.2.1). We expect this ring lead to systematic effects on the earlier-published IOTA astrometry. By re-fitting the IOTA data, taking the extended flux into account, we obtain binary separations for the inner (A-B) system that are 21 to 42% smaller than those used for the earlier orbit solution (*14*). The separations for the outer component are less affected.

The model fitting was conducted using the ORBIT3 code (*61, 62*); we fit the astrometry points (Table S4) simultaneously with the radial velocity data (*15*).

The best-fit orbit solution is presented in Table S5.

Both orbits are retrograde, i.e. the companions move in the direction of decreasing PA, with the ascending node towards the South-West (Fig. S8). Comparing our solution to the previously published solution (*15*) (Fig. S9), we find that the short-period orbit is less eccentric (0.069 versus 0.13), while the long-period orbit has higher eccentricity that previously assumed (0.379 versus 0.13 or 0.25 for the two previous solutions). Our orbital elements are consistent with the values for the A-B binary orbit derived indepedently (*13*).

We compute the mutual inclination $\Phi$ between the two orbit planes from their angular momentum vectors (*63*):

$$\cos\Phi = \cos i_{\mathrm{A-B}} \cos i_{\mathrm{AB-C}} + \sin i_{\mathrm{A-B}} \sin i_{\mathrm{AB-C}} \cos(\Omega_{\mathrm{AB-C}} - \Omega_{\mathrm{A-B}}), \quad (\mathrm{S1})$$

where $i_{\mathrm{A-B}}$ and $i_{\mathrm{AB-C}}$ are the inclination of the inner and outer system and $\Omega_{\mathrm{A-B}}$ and $\Omega_{\mathrm{AB-C}}$ are the position angle of the ascending node of the inner and outer orbit. We compute that the A-B and AB-C orbital planes are misaligned by $\Phi = 13.9 \pm 1.1°$. The 3-dimensional orientation of the orbits is illustrated in Fig. S10 while Fig. S11 shows the on-sky-projected orbits overplotted on VLTI/AMBER and MIRC-X aperture synthesis images.

### S1.2.3 Dynamical masses and orbital parallax

Earlier studies adopted a broad range of mass estimates for GW Ori. For instance, (*64*) estimated the mass using evolutionary tracks, yielding for GW Ori A a mass $M_A = 2.5$ M$_\odot$ and for GW Ori B $M_B = 0.5$ M$_\odot$, while (*14*) derived masses of $M_A = 3.6$ M$_\odot$ and $M_B = 3.1$ M$_\odot$. However, most of these earlier estimates were based on measurements of the infrared flux ratio between the primary and secondary. Our observations show that this flux ratio is highly variable (Table S4), possibly due to a combination of variable extinction and phase-dependent thermal emission contributions from circumstellar dust. Therefore, we consider the infrared flux ratios as unsuitable to constrain the stellar masses in the system.

The masses of the inner binary (A+B) and the total system (A+B+C) are derived from the measured semi-major axes and periods using Kepler's third law, yielding $M_{A+B} = 3.90 \pm 0.40$ $M_\odot$ and $M_{A+B+C} = 5.26 \pm 0.22$ $M_\odot$, respectively. Furthermore, the $M_A \sin^3 i_{A-B}$ mass function for A and B are constrained from the measured radial velocity semi-amplitudes, eccentricity, and orbital period. Correcting for the measured inclination yields $M_A = 2.47 \pm 0.33$ $M_\odot$ and $M_B = 1.43 \pm 0.18$ $M_\odot$. The radial velocity of the tertiary itself remains unmeasured – therefore we can only derive a minimum mass for C from the mass function, $M_C > 0.78 \pm 0.10$ $M_\odot$. However, based on the total mass constraints, we find that the mass of GW Ori C is $M_C = M_{A+B+C} - M_{A+B} = 1.36 \pm 0.28$ $M_\odot$. Our uncertainty estimates include the distance uncertainties as listed in Table S3.

By combining the linear separation from the spectroscopic orbit and the angular separation measured with interferometry, we can derive the orbital parallax of the system and determine the distance to GW Ori. We compute the linear scale for the A-B system in astronomical units (*65*)

$$a_A = 9.1913 \times 10^5 \frac{P \cdot K_A \cdot \sqrt{1-e^2}}{\sin(i)} \tag{S2}$$

$$a_B = 9.1913 \times 10^5 \frac{P \cdot K_B \cdot \sqrt{1-e^2}}{\sin(i)}, \tag{S3}$$

where the period $P$ is given in days, $K_A$ and $K_B$ are in km s$^{-1}$. The parallax is then $\pi = a/(a_A + a_B)$, where $a$ is the angular semi-major axis in arcseconds. Using the orbital elements from Table S5, this yields a distance of $387 \pm 27$ pc, which is consistent with the distance of $388 \pm 5$ pc reported for the Orion Nebula Cluster (*11*) and the $398 \pm 10$ pc from the Gaia DR2 parallax for GW Ori (*66, 67*).

### S1.2.4 Disk structure: sub-millimeter thermal emission

To quantify the ALMA data, we model the observed brightness distribution using a geometric model. In the model we include the three bright rings labeled R1, R2 and R3 in Fig. 1A. In addition, we include a fourth, broad ring that represents the extended emission located between the rings, in the following referred to as $R_{\mathrm{disk}}$. Our model includes the following components, where any offsets are measured with respect to the visible-light astrometric position of GW Ori:

i) three rings ($j = 1, 2, \mathrm{disk}$) with radii $r_{\mathrm{Rj}}$ and a Gaussian radial intensity profile with half-width-at-half-maximum (HWHM) $\Theta_{\mathrm{Rj}}$ and flux density $F_{\nu,\mathrm{Rj}}$ to represent rings R1, R2, and $R_{\mathrm{disk}}$. The rings are projected to mimic inclination effects (with inclination angle $i_{\mathrm{Rj}}$ and the PA of the ascending node $\Omega_{\mathrm{Rj}}$), where the inclination and PA for $R_{\mathrm{disk}}$ are fixed to the values for R1, i.e. $i_{\mathrm{Rdisk}} = i_{\mathrm{R1}}$ and $\Omega_{\mathrm{Rdisk}} = \Omega_{\mathrm{R1}}$. To define the ascending node, we assume that the rings rotate in retrograde motion (i.e. in clockwise direction on the sky), following the same rotation direction as the stellar orbits. We tested whether the fit improved if these components were offset with respect to the origin of the coordinate system – however, there was no significant improvement in the fit and therefore we fixed the centers on the position of the stars.

ii) another ring, representing ring R3, with the same free parameters, but an additional azimuthal modulation that we parameterize as $f_{\mathrm{R3}}(\theta) = (1 - a_{\mathrm{R3}} \sin(\theta - \theta_{\mathrm{R3asym}}))^{\gamma_{\mathrm{R3asym}}}$. $a_{\mathrm{R3}}$ is the amplitude of the asymmetry, $\gamma_{\mathrm{R3asym}}$ the stretch factor, $\theta_{\mathrm{R3asym}}$ the azimuth angle of asymmetry, and $\theta$ the azimuth angle. The ring is allowed an offset with respect to the origin of the coordinate system ($\Delta\alpha_{\mathrm{R3}}$, $\Delta\delta_{\mathrm{R3}}$).

iii) two Gaussians with HWHM $\Theta_{\mathrm{AB}}$ and $\Theta_{\mathrm{C}}$ and flux density $F_{\nu,\mathrm{AB}}$ and $F_{\nu,\mathrm{C}}$ to represent the inner components $D_{\mathrm{AB}}$ and $D_{\mathrm{C}}$. The location of the first component is fixed at the origin of the coordinate system, while the offset of the second Gaussian are free parameters ($\Delta\alpha_{\mathrm{C}}$, $\Delta\delta_{\mathrm{C}}$).

The geometric model is convolved with the interferometric beam and then fitted to the observed image with a Differential Evolution optimisation algorithm (*68*). The best-fitting pa-rameters are reported in Table S6, where the reported uncertainties have been estimated from the $\chi^2$ surface near its minimum. The model parameters for R1, R2, R3, and R$_{\rm disk}$ show a clear global minimum. The values derived for the ring components assumes that the underlying 3-dimensional geometries are circles, which might not be justified for ring R3. Therefore, in Sect. S1.2.5 we will consider that the R3 geometry might be an ellipse seen under intermediate inclination, which allows us also to reproduce the shadows seen with polarimetric imaging. The Gaussians D$_{\rm AB}$ and D$_{\rm AB}$ have a weak correlation between the HWHM ($\Theta_{\rm AB}$, $\Theta_{\rm C}$) and the offset between the components ($\Delta\alpha_{\rm C}$, $\Delta\delta_{\rm C}$) once the HWHM becomes larger than the separation between the components. We avoid these degenacies by forcing the components to be unresolved ($\lesssim 4\,\rm mas$).

Using equation S1, we compute the mutual inclination between the orbits and the disk planes. We find that the angular momentum vector between the inner binary orbit (A-B) and the outer-most ring R1 are misaligned by $51.1 \pm 1.1°$. For the outer orbit AB-C, the mutual misalignment is $38.5 \pm 0.8°$. The 3-dimensional orientation of the disk planes is illustrated in Fig. S10.

We estimate the dust mass in the rings from the sub-millimeter flux density. Assuming that the emission is optically thin,

$$M_{\rm dust} = \frac{d^2 F_\nu}{B_\nu(T_d)\, \kappa_\nu}, \tag{S4}$$

where $F_\nu$ is the sub-millimeter flux density measured with ALMA. $T_d$ the dust temperature, and $B_\nu(T_d)$ the blackbody function. $\kappa_\nu$ is the dust opacity per dust mass, for which we adopt $\kappa_\nu = 0.2(7\,\rm mm/\lambda)\,cm^2 g^{-1}$ (*69, 70*).

For a disk in hydrostatic equilibrium, (*71*) showed that the gas temperature (which can be used as proxy for the dust temperature in the thermally coupled case) can be estimated from

$$\frac{c_s}{R\Omega_a} = H_p/R \tag{S5}$$

$$c_s^2 \approx R_g T_d/\mu, \tag{S6}$$

where $c_s$ is the sound speed, $R_g$ is the universal gas constant, $\mu$ the mean molecular weight. $H_p/R$ is the pressure scale height of the disk at radius $R$, which we set to 0.05 based on the hydrodynamic simulation (see below) and the sharpness of the radial shadows (S1+S2) that we see in the scattered light images. $\Omega_a$ denotes the azimuthal frequency $\Omega_a = \sqrt{GM/R^3}$. For the location of $D_{AB}$ and $D_C$, R3 and $R_{disk}$, R2, and R1, we estimate the dust temperatures to 650 K, 71 K, 23 K, and 18 K, respectively. The resulting dust mass estimates are reported in Table S6. The sum of mass of the three rings R1, R2, and R3 is $0.13\,M_\odot$ or $0.21\,M_\odot$ if extended dust emission in $R_{disk}$ is included (combined gas+dust, assuming a gas-to-dust ratio of 100, (*72*)). These estimates are consistent with the $0.12\,M_\odot$ derived previously (*45*).

### S1.2.5  Disk structure: visible/near-infrared polarimetry

Our four SPHERE and GPI coronagraphic-polarimetric images reveal scattered light originating from the disk surface. The polarized flux is $\sim 4$-times higher towards the East of the stars than towards the West. Given that polarized intensity images are typically dominated by forward-scattering from dust grains, this indicates that the Eastern side of the disk is facing towards the observer. This conclusion is also supported by CO rotation measurements (*15, 45*) that show that the Northern part of the disk is receeding from the observer (red-shifted). If the disk rotates in retrograde motion (i.e. in clockwise direction on the sky), equivalent to the stellar orbits, then the Eastern side must face towards us.

The emission is inhomogeneous and appears to be arranged broadly in four arcs (A1, A2, A3, and A4; labeled in Fig. 1C), where the drop in polarized intensity between the arcs coincides with the position of the sub-millimeter rings R1, R2 (Figs. 1E and S3, bottom row). We interpret this as a shadowing effect: The rings R1 and R2 might appear bright in sub-millimeter continuum emission due to trapping of mm-sized grains near a pressure maximum, similar to what has been proposed for continuum rings seen in circumstellar disks (e.g. (*73*)). In this case, we expect an enhanced disk scale height at the radial locations of the sub-millimeter rings, which results in shadowing from the stellar photons in the regions immediately behind the pressure bump (Fig. S13). Therefore, these regions appear dimmer in scattered light images.

We see four radial shadows, including two sharply defined shadows extending in South-East and North-West direction (S1 and S2) and two broader shadows extending to the North-East and South-West (S3 and S4). To understand the origin of these shadows, we constructed a 3-dimensional scattered light model using the ray-tracing software package BLENDER (version 2.79c; (*74*)). We parameterize geometrically thin disk fragments that resemble ALMA ring R1, ring R2, a coplanar component connecting R1 and R2, and a surface that connects R2 with an inner truncation orbit (which we refer to as the 'break orbit'). In addition, we include an eccentric, vertically extended ($H_p/R = 10\%$), and optically thick (i.e. opaque) ring that resembles ALMA ring R3 (Table S6), allowing sharp shadows to be cast on the outer disk.

Following the morphologies seen in hydrodynamic simulations (e.g. (*3,4*)), we model the disk surface between R2 and the break orbit as warped dust surfaces. These surfaces is build-up by connecting a set of titled rings with radii $r_n$. We parameterize the radial profile of these rings as a Fermi function in polar coordinates, where the rings follow the slope of the Fermi function $\sim 1/(e^{-r_n} + 1)$. This results in a smooth transition between rings of different inclination. Surfaces are defined by connecting the longitudes on an inner boundary orbit with the equivalent longitudes on an outer boundary orbit. These boundary orbits are allowed to have different values for inclination, eccentricity and longitude of the ascending node. The surfaces are modeled as geometrically thin scattering surfaces with intermediate optical depth, i.e. they permit some light to propagate through. The optical properties are only used to simulate shadow morphologies; we do not aim to reproduce the scattered light images quantitatively.

The free parameters to define the 3-dimensional shape and orientation of R3 are the inclina-tion $i_{R3}$, eccentricity $e_{R3}$, and argument of periastron $\Omega_{R3}$. To match the shadow pattern (seen in the SPHERE and GPI images) simultaneously with the on-sky projected shape of the occulting ring R3 (seen in the ALMA image), we fix $a_{R3} = b_{R3}/\sqrt{1-e_{R3}^2}$ and $\omega_{R3}$ based on $i_{R3}$, $e_{R3}$, and $\Omega_{R3}$.

We place three light sources in the center of the model at positions computed from the GW Ori orbit solution. Scattering from the disk surface is computed through ray-tracing in BLENDER. We convolve the resulting synthetic images to match the resolution of the SPHERE/GPI observations and adjust the free model parameters to obtain the best match on the direction and shape of the shadow patterns between the model and our scattered light imagery (Figs. 1C+D). We reproduce the shape of the apparent ellipse that is formed by the scattered light arcs A3 and A4 (Fig. 1C). The orientation of this apparent ellipse (with semi-major axis oriented along PA$\sim 60 \pm 10°$) differs notably from the orientation of the outer disk ($\Omega_{R2} = 180 \pm 4°$). Our scattered light model allows us to reproduce this shape approximately by choosing the following parameters for the break orbit: $a_{break} = 90$ au, $e_{break} = 0.65$, and $\Omega_{break} = 60°$. Choosing an inclination value $i_{break} = 15°$ allows us to reproduce the prominent arc A3 as scattered light from the side of the warped disk that is facing away from us and that is directly illuminated by the stars (Fig. S13). The much dimmer arc A4 corresponds to light from the warp surface that is facing towards us, where we see only the self-shadowed outer side of the warp that is not directly illumined. The highest column density in the warp is towards the North-East and South-West, matching the directions where we see the broad shadows S3 and S4 (Fig. 2A).

The derived parameters for ring R3 are listed in Table S6 and show that ring R3 is eccentric ($e_{R3} = 0.3 \pm 0.1$). This is consistent with the off-center position of the ring with respect to the stars, if one of the focal points of the ellipse coincides with the center-of-mass of the system. As the ring is strongly inclined with respect to the plane of the sky, it casts a narrow radial shadow, both in South-East direction (matching S1) and North-West direction (matching S2). The model reproduces the radial curvature seen in shadow S1 as a geometric effect, where the shadow cast by R3 is projected onto the warped surface. The inner part of the shadow (S1$_{\text{inner}}$) is projected on the warped surface, while the outer part (S1$_{\text{outer}}$) is projected on the non-warped surface connecting R1 and R2.

## S1.3 Hydrodynamic modeling

Using our constraints on the 3D orbits and stellar masses, we ran SPH simulations of the GW Ori system. Our simulations were performed using the SPHNG code (*75*), (*76*), and (*77*), which has previously been applied to higher-order multiple systems (e.g. (*76*)) and protoplanetary disks (e.g. (*78*)). Some figures were produced using the SPH visualization tool SPLASH (*79*). The simulation setup is based on our measured stellar orbits and the GW Ori stellar and disk parameters listed in Table S3.

To initialize the sink particles in the SPH simulation we calculate the initial positions and velocities at JD 2456674.8 (2014 January 17), i.e. the time when GW Ori B passes through periastron and has a true anomaly of $0°$. The periastron passage of GW Ori C occured 2815.2 days earlier so we solve the Kepler equation to obtain the true anomaly of GW Ori C 2815.2 days after periastron passage.

In the SPH models we define the positive $\hat{x}$ and $\hat{y}$ directions as North and East respectively. The $\hat{z}$ axis points towards the observer. We perform a rotation to position the stars in the model coordinates. The orbital elements $\Omega$, $i$ and $\omega$ loosely correspond to the Euler angles $\alpha$, $\beta$ and $\gamma$, but need to be adjusted to account for differences in their definitions.

The longitude of the ascending node, $\Omega$, is measured anti-clockwise from North to the point where the orbital plane intersects the plane of the sky and the motion of the secondary is away from the observer, into the sky. In mathematical convention, the ascending node is defined as the node where the motion has a positive $\hat{z}$ component. This is the opposite node to the astronomical convention so we set $\alpha = \Omega - \pi$ radians. The inclinations listed in Table S6 are measured at the ascending node, clockwise looking down the ascending node. Because $\beta$ is measured anticlockwise around the opposite node, $\beta = i$. In Table S5, the visual binary conventions are used for defining $\omega$. Accordingly, $\omega$ is measured in the direction of motion of the companion from the ascending node of the secondary and so $\gamma = \omega$. We then apply standard Euler rotations about $z - x' - z''$, with negative rotation angles, in the order $-\gamma$, $-\beta$, $-\alpha$, moving from the frame of the orbit to the frame of the sky.

The system is considered as a hierarchical triple with motion of A and B relative to their mutual center-of-mass; the motion of C and the A-B center-of-mass are treated separately. The positions are calculated in the frame of the orbits from the true anomaly and the barycenter distance found from the stellar masses. The position vectors are then rotated to the sky orientation as described above and the positions of A & B are shifted relative to the A-B center-of-mass calculated from the orbit of C. The relative velocities are calculated, similarly, first in the frame of the orbits according to the standard formulae for an elliptical orbit and then scaled for the component mass. The velocity vectors are then rotated onto the frame of the sky and the velocity of the A-B center-of-mass due to C is added to the velocity of A and B. The disk is positioned by following a similar rotation but excluding the $-\gamma$ rotation since it is initially axisymmetric.

To model the disk we use $8 \times 10^5$ SPH gas particles. We begin with a disk that has a 200 au outer radius and a 20 au inner radius. The surface density profile $\Sigma(R)$ follows $\Sigma(R) \propto R^{-0.2}$ (*45*), with a scale-height given by a fixed $H_p/R = 0.02$. We do not model the outer part of the observed disk because we are studying the warping and/or disk tearing of the inner disk and only modeling the inner 200 au reduces the computational expense. We set up the SPH particles in a Keplerian disk orbiting a single gravitational mass of 5.26 M$_\odot$ (i.e. the total stellar mass of the system), and the disk is evolved until any transient structures from the initial conditions have dissipated. We then add the three stars as point masses (sink particles with accretion radii of 1 au), and reorientate the settled disk to match the orientation inferred from observations for the outer disk (ring R1). We also remove any disk material within 40 au of the center-of-mass of the stellar system. This gives the initial conditions for our hydrodynamical simulation. We run the simulation without self-gravity of the gas, and exclude the gravitational force from the disk on the stars so that the orbital parameters of the stellar system stay close to the observed values throughout the simulation.

We use an SPH artificial viscosity parameterized with $\alpha_{\rm SPH}$ and $\beta_{\rm SPH}$ (*80*), with $\alpha_{\rm SPH}$ ranging between 0.1 and 1, and $\beta_{\rm SPH}$ fixed at 2. The $\alpha_{\rm SPH}$ viscosity provides a viscosity that has a similar effect to a Shakura & Sunyaev viscosity $\alpha_{\rm SS}$ (*71, 81*). In our simulation the magnitude of the equivalent Shakura & Sunyaev viscosity is $\alpha_{\rm SS} \approx 0.01 - 0.02$.

Once the hydrodynamical simulation is started, the inner region of the disk slowly develops a warp over the first 1000–3000 years and the inner edge spreads inward. At around 5500 years an inner ring starts to develop and detaches from the rest of the disk. Its initial radius is approximately 30 au. It precesses rapidly and from about 6500 years onward it is distinct from the rest of the disk. Over the next 3000 years, the precessing ring occasionally interacts with the inner edge of the warped outer disk, accreting gas, growing in radius, and becoming more eccentric. Similar behavior has been observed in other simulations for broken disks (*5*). At 9500 years, shown in Fig. 3, the inner ring has a radius of approximately 40 au and an eccentricity of $\sim 0.2$, and the outer part of the disk is warped. These structures are in qualitative agreement with those that are observed for the GW Orionis system.

We find that the simulation displays the main features that we have used to model and explain the observations including disk tearing, the formation of an inner, precessing, eccentric ring that has similar dimensions to the observed inner ring, and a warped outer disk. Differential precession naturally occurs in the hydrodynamical model, preventing an exact match with the observations. Furthermore, parameters such as the exact scale-height and mass of the disk and its viscosity are unknown.

We interpret rings R1 and R2 as dust traps triggered by the strong density gradient inside of R2. Besides dust trapping at pressure maxima, dust pile-up could alternatively occur at the location of disk warps due to a difference in precession between the gas and dust (*26*). However, this mechanism applies only in the high Stokes number (St $\gtrsim 10$) regime (*26*). We estimate the Stokes number for the dust grains traced by our ALMA images based on the initial surface density profile listed above and assuming a particle internal density of $1\,\mathrm{g\,cm^{-3}}$. We find that the emission seen in our ALMA image is likely dominated by particles with sizes of $\lambda/2\pi \approx 0.1\,\mathrm{mm}$ with a typical Stokes number between $\sim 0.001$ (at 30 au) and $0.0035$ (at 200 au). For this regime (St $\lesssim 0.1$) the dust should closely follow the gas, indicating that the differential precession dust trapping mechanism should not affect the morphologies seen in our ALMA image (*26*).

## S2 Supplementary Text

### S2.1 Analytic break radius estimate

The strongly misaligned eccentric ring R3 seen in our sub-millimeter imaging shows many characteristics that are predicted for disk tearing. For a quantitative comparison with earlier theoretical predictions, we compare the measured R3 radius to analytic estimates of the break radius $r_\mathrm{break}$ that is defined as the point where the external torque exerted by the misaligned binary exceeds the internal torque due to pressure forces (*4, 5*):

$$r_\mathrm{break} \lesssim 50\mu_C^{1/2} |\sin 2\Phi|^{1/2} \left(\frac{H_p/R}{0.1}\right)^{-1/2} \left(\frac{\alpha_\mathrm{SS}}{10^{-3}}\right)^{-1/2} a, \quad (\mathrm{S7})$$

where $\mu_C = M_C/(M_A + M_B + M_C)$ is the mass fraction of the tertiary, $\Phi$ is the initial misalignment between the disk and the binary ($\Phi = 51.1 \pm 1.1°$ for the AB-C system; Sect. S1.2.4), $H_p/R$ is the pressure scale height of the disk, and $\alpha_{\rm SS}$ the viscosity.

As this equation was derived for misaligned binary systems (*4, 5*), it has to be applied with caution to a triple system such as GW Ori. However, we expect the torque on the outer disk to be dominated by the wide-separation tertiary and therefore apply the estimate to the AB-C system. Adopting $H_p/R = 0.05$ (Sect. S1.2.4), we find that the break radius estimate matches the size of the R3 semi-minor axis ($b_{\rm R3} = 43$ au) for $\alpha_{\rm SS} \lesssim 0.05$ (or any case where the product $\alpha \cdot H_p/R \lesssim 0.0025$). As most estimates for the $\alpha_{\rm SS}$ viscosity in protoplanetary disks range between 0.001 and 0.04 (*82*), we conclude that the disk around GW Ori is susceptible to disk tearing at the location of ring R3.

## S2.2  Carbon Monoxide map interpretation

The Carbon Monoxide spectral line map (zeroth-moment map; Fig. S1A) shows that the CO surface brightness is highest in the warped disk region between continuum rings R2 and R3 (labeled C1 in Fig. S1A) and at the inner edge of ring R3 (labeled C2). The ring R3 itself appears as a region of low CO surface brightness. We interpret this as a gas temperature effect, where the high dust opacity in the ring R3 shields CO gas located within the ring from direct stellar illumination. Due to the viewing geometry (Sect. S1.2.5 and Fig. S13), we see on the Eastern side of the ring R3 the inner surface that faces towards the stars; on the Western side we look onto the outer ring surface. The CO gas at the inner ring surface is directly illuminated by the star and has therefore a higher gas temperature, which translates into a higher CO surface brightness. This matches the observation, where the Eastern side of C2 exhibits a higher CO surface brightness than the Western side. Similarly, the warped disk region located South-West of the stars is directly illuminated by the stars, which explains the high CO surface brightness seen in the C1 region.

The first-moment map (Fig. S1B) shows the intensity-weighted line-of-sight velocity of the Doppler-shifted CO gas. The velocity gradient seen on the largest scales ($> 200$ au) is broadly consistent with the one seen in the outer disk (*15, 45*), with the Northern part of the disk receeding from the observer (red-shifted). In the inner 200 au we see a twist in the first-moment map, whose morphology is consistent with a pattern seen previously in a lower-resolution ALMA observation (*83*). We find that the twist forms a spiral arm pattern, where the position angle of the axis between the receeding/approaching gas motion (red-shifted/blue-shifted line emission) shifts from North-South direction (PA$\sim 0°$ at 200 au separation from the star) to East-West direction (PA$\sim 90°$ at 100 au) and to South-East/North-West direction (PA$\sim 120°$) inside of ring R3 ($\lesssim 30$ au). These gas motions connect the outer disk with the inner disk and facilitate the accretion on the circumbinary disk $D_{AB}$ and circumtertiary disk $D_C$, and onto the stars.

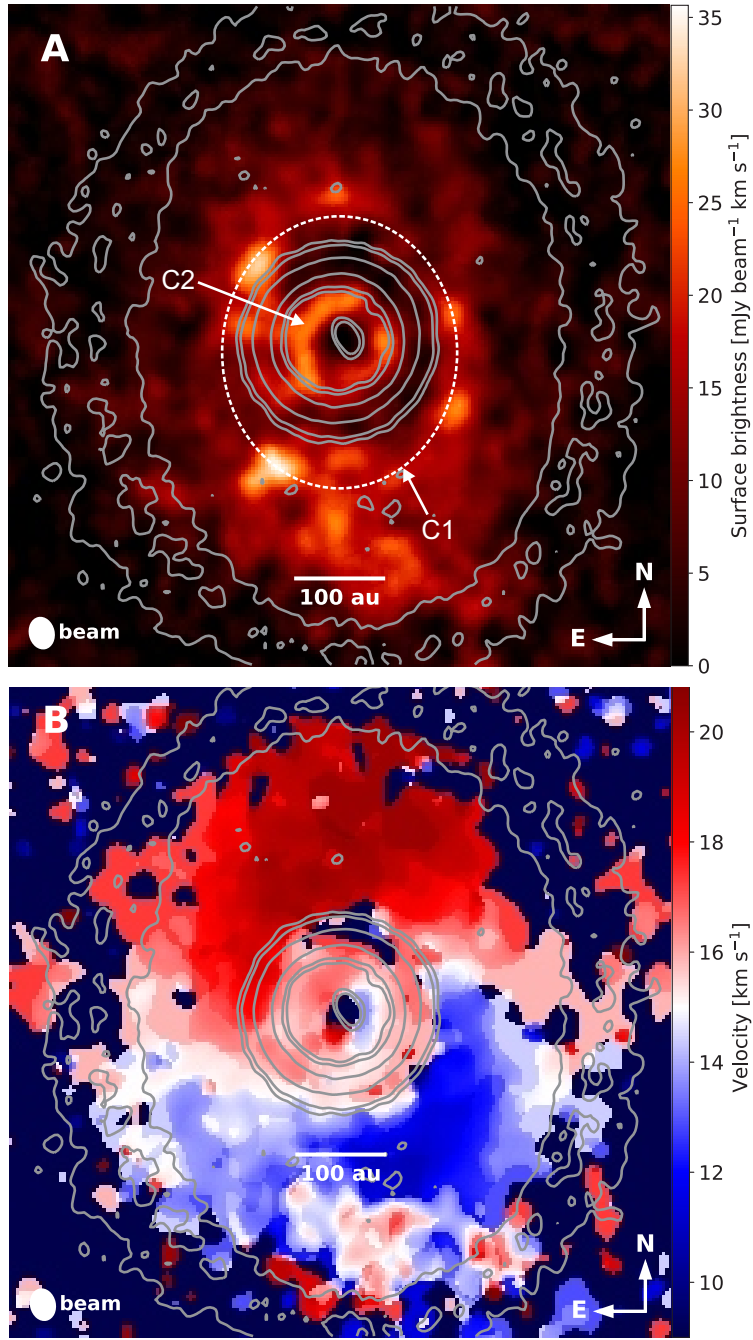

Figure S1 – **Carbon Monoxide spectral line maps of GW Ori.** (A) Surface brightness (zeroth-moment) map in the $^{12}$CO line. (B) Intensity-weighted velocity (first-moment) map, where velocities are measured with respect to the local standard of rest kinematic (LSRK) system that is based on the average velocity of stars in the solar neighborhood. In both panels, we show the interferometric beam and overplot the contours of the 1.3 mm continuum images at levels of 3.5%, 10%, and 50% of peak intensity. The beam size and on-sky orientation shown in panel A applies to both panels.

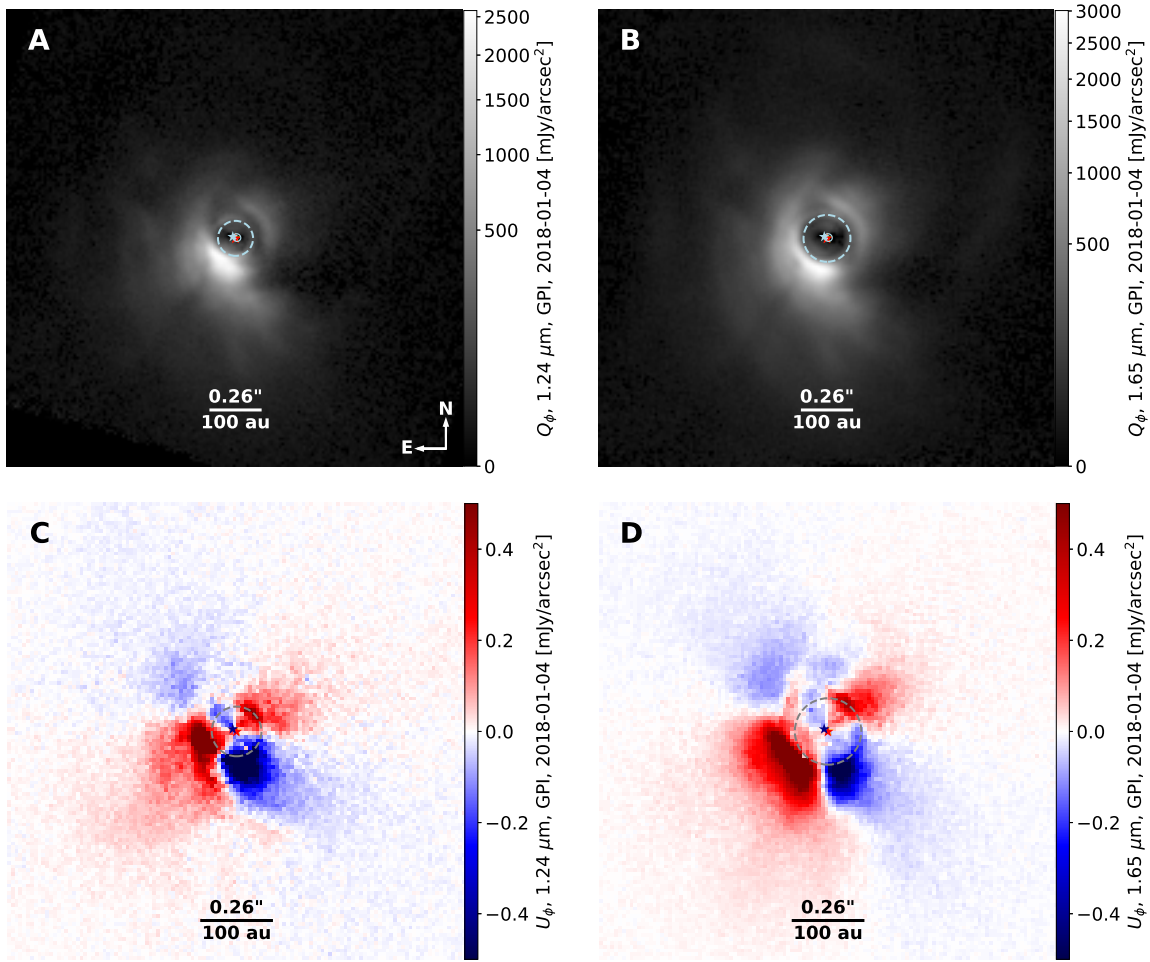

Figure S2 – **Polarized intensity images of GW Ori.** The images show Stokes $Q_\phi$ (A and B) and $U_\phi$ (C and D) retrieved with GPI in J-band (A and C) and H-band (B and D). The dashed circle indicates the inner-working angle of these coronagraphic imagng observations. We also show the position of the three stars (star symbols) at the time of the observation and, in (A) and (B) the triple star orbits (solid curves). In all images North to up and East to the left, as indicated by the compass in (A).

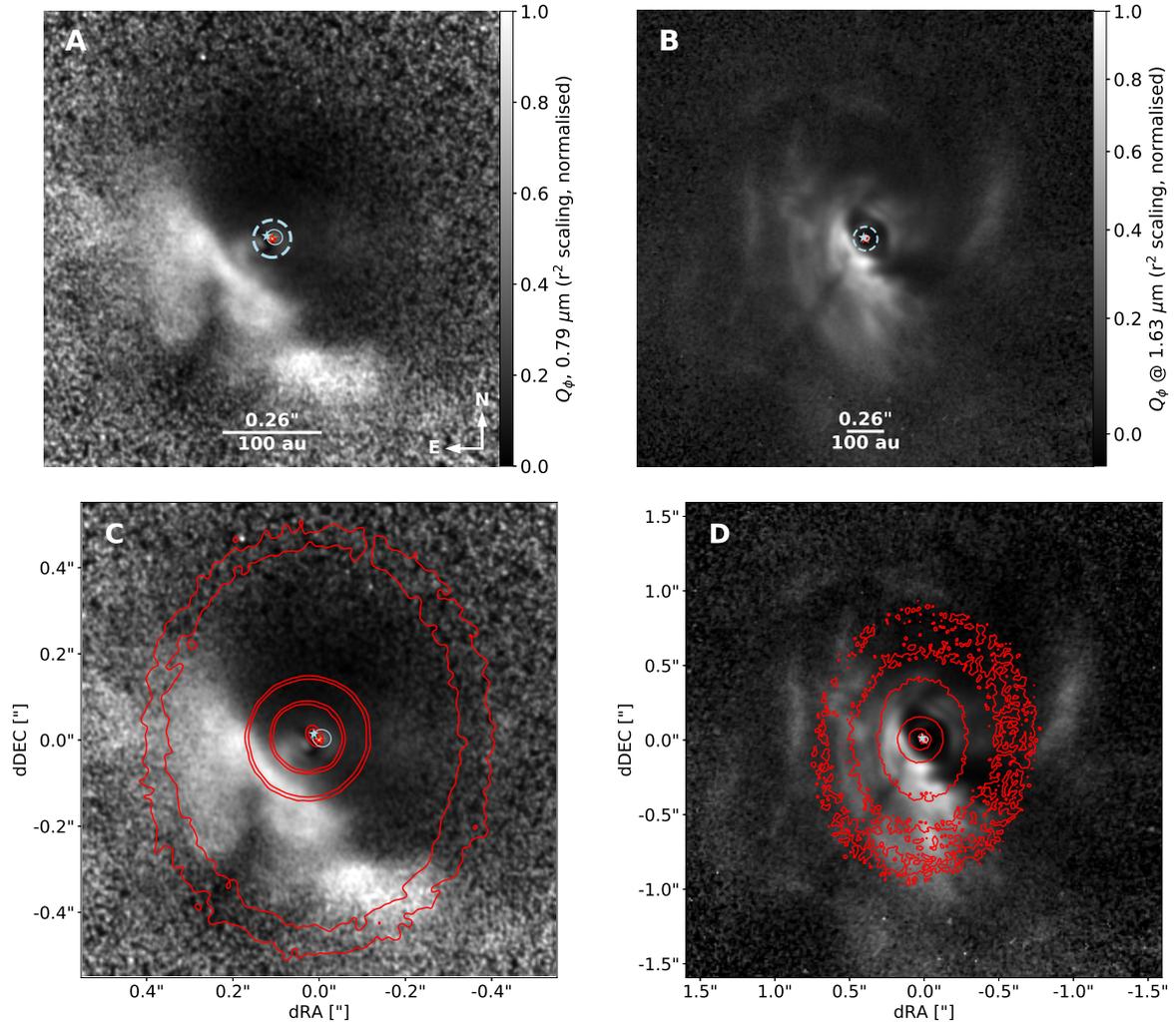

Figure S3 – **Polarized intensity $Q_\phi$ images of GW Ori recorded with SPHERE/ZIMPOL (A and C) and SPHERE/IRDIS (B and D).** The images have been multiplied with $r^2$ to emphasize structures in the outer disk, where $r$ is the distance from the star. The dashed blue circle indicates the inner working angle, marking the region where the imaged structures are affected by residual star light contributions (for ZIMPOL) or the coronagraphic mask (for IRDIS). In (C) and (D), we overlay the ALMA data as red contours at levels of 15% (C) and 8% (D) of peak intensity. As in Figure S2, we also indicate the position of the stars at the epoch of the observation (as derived from our orbit solution) and the orbit of the tertiary component.

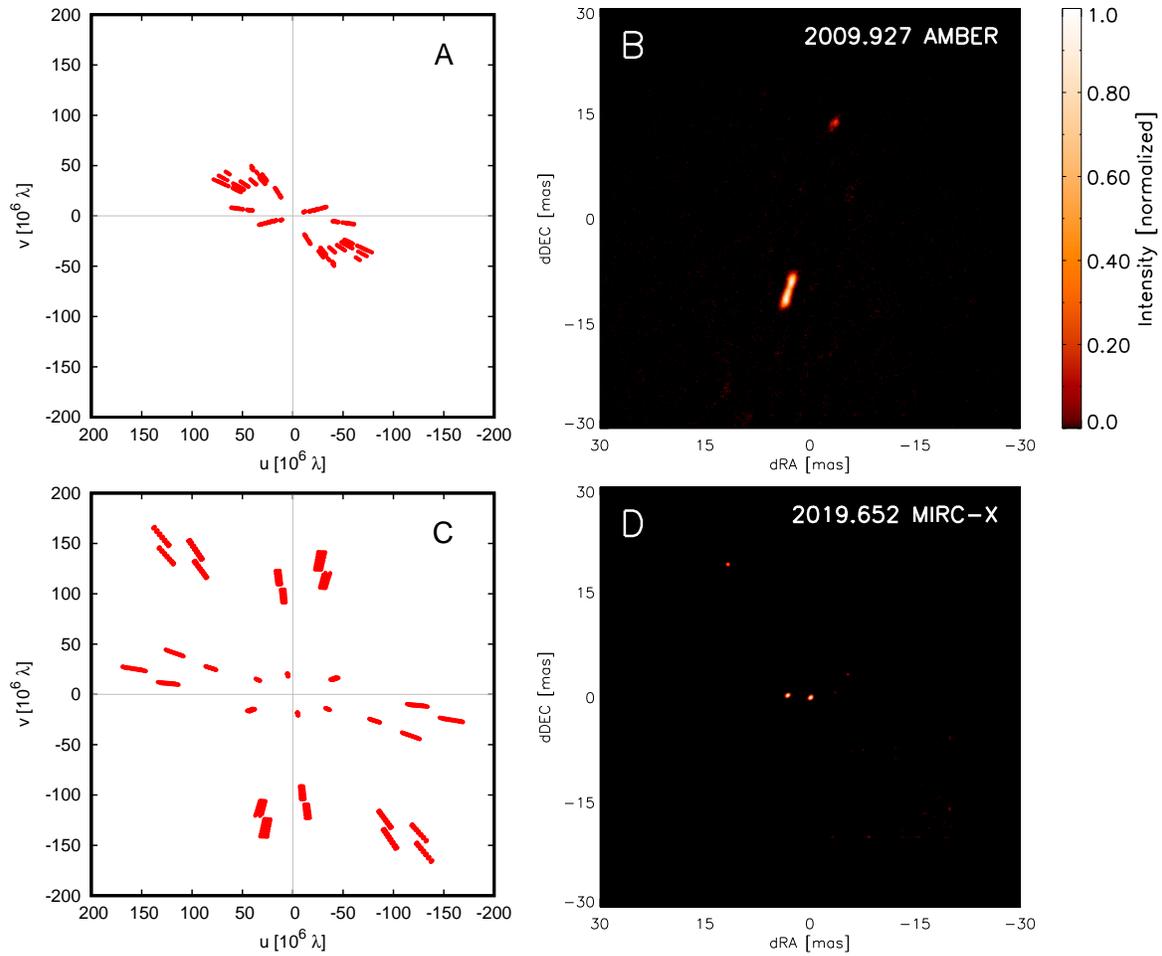

Figure S4 – **Near-infrared interferometric images of the GW Ori triple system obtained with AMBER on 2009 December 4 (A and B) and MIRC-X on 2019 August 27 (C and D).** (A and C): $(u,v)$-coverage. (B and D) corresponding aperture synthesis images.

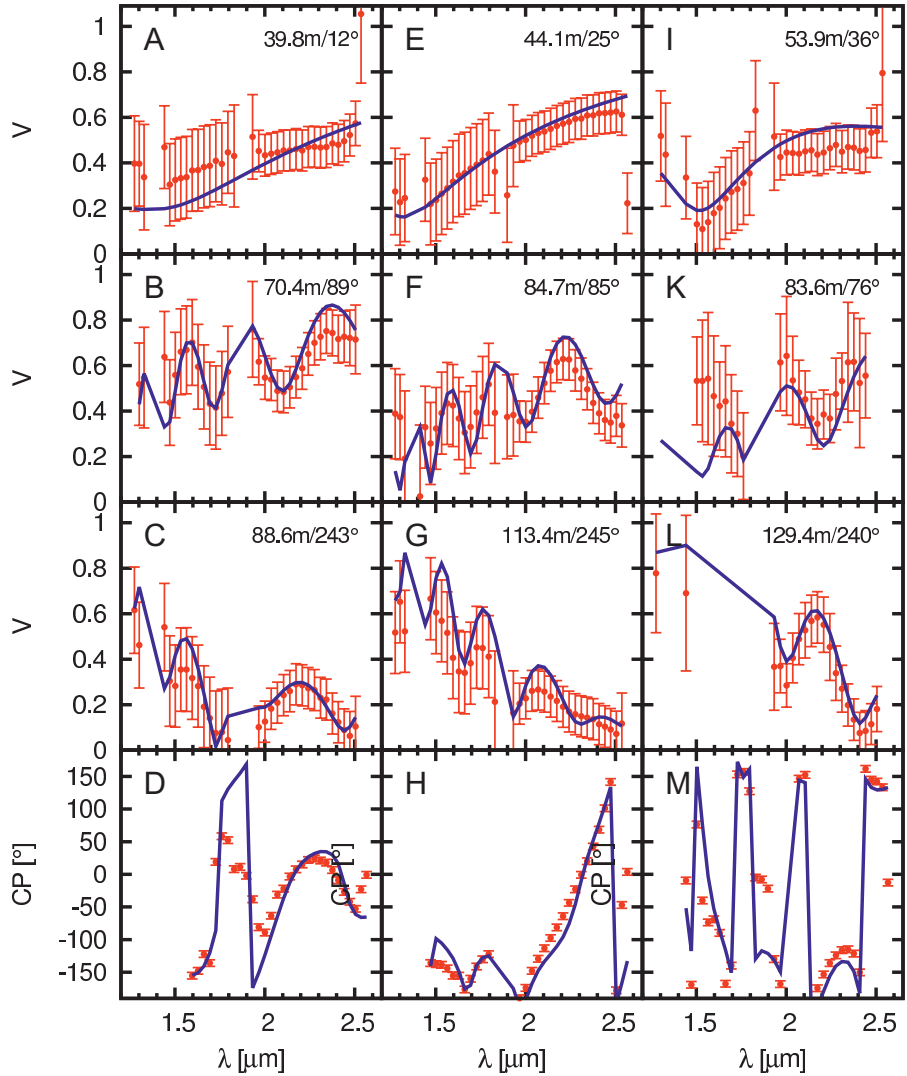

Figure S5 – **Interferometric observables recorded on GW Ori with VLTI/AMBER on 2012 October 29.** Data recorded on GW Ori at three hour angles (A-D: UT 05:20, E-H: UT 06:38, I-M: 09:15) with the telescope configuration UT1-UT2-UT4. V indicates visibility amplitudes, where the length and PA of the sky-projected baseline vector is noted in the label of each panel. CP indicates the closure phases that have been measured on the closed triangle formed by the three baseline vectors of each measurement. We over-plot the triple star + smooth ring model outlined in Sect. S1.2.1 (blue curves). The errorbars indicate $1\sigma$ (95%) uncertainties and include both statistical and calibration uncertainties.

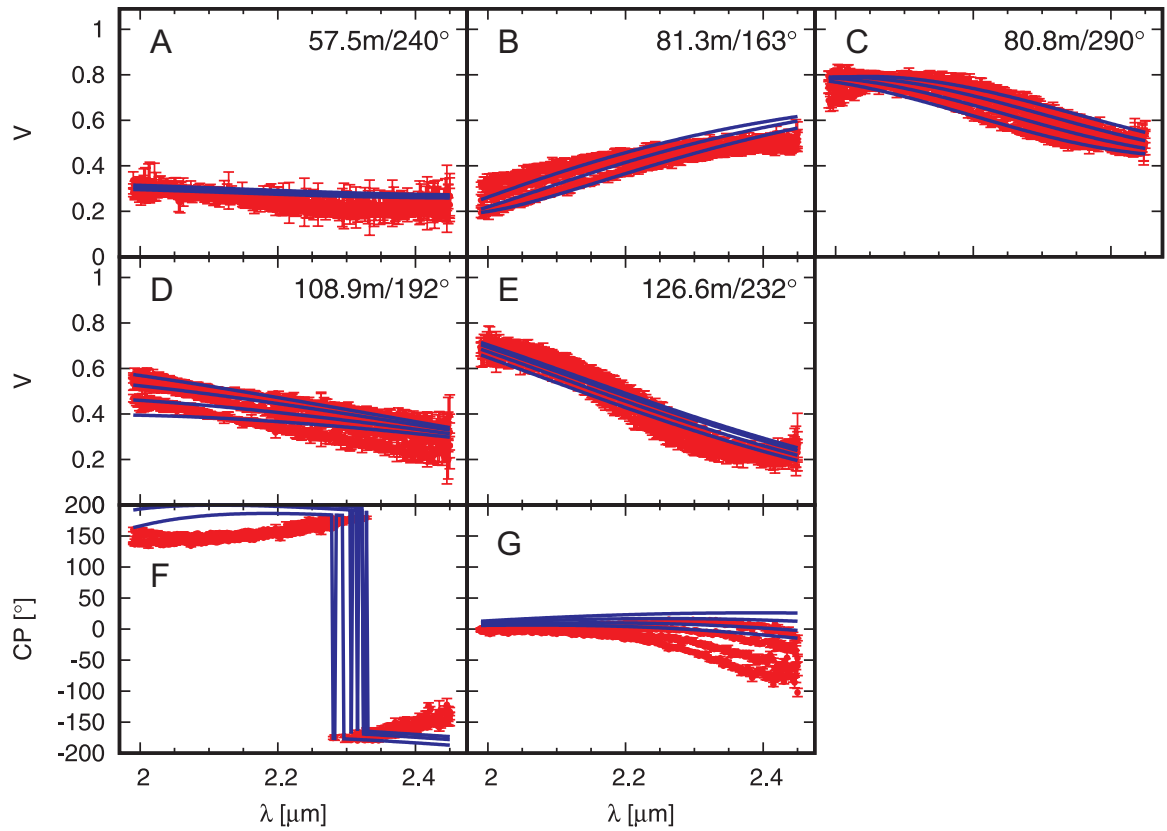

Figure S6 – **Same as Fig. S5, but for the interferometric observables recorded with VLTI/Gravity on 2018 February 5.** This data has been recorded with the telescope configuration A0-G1-J2-J3.

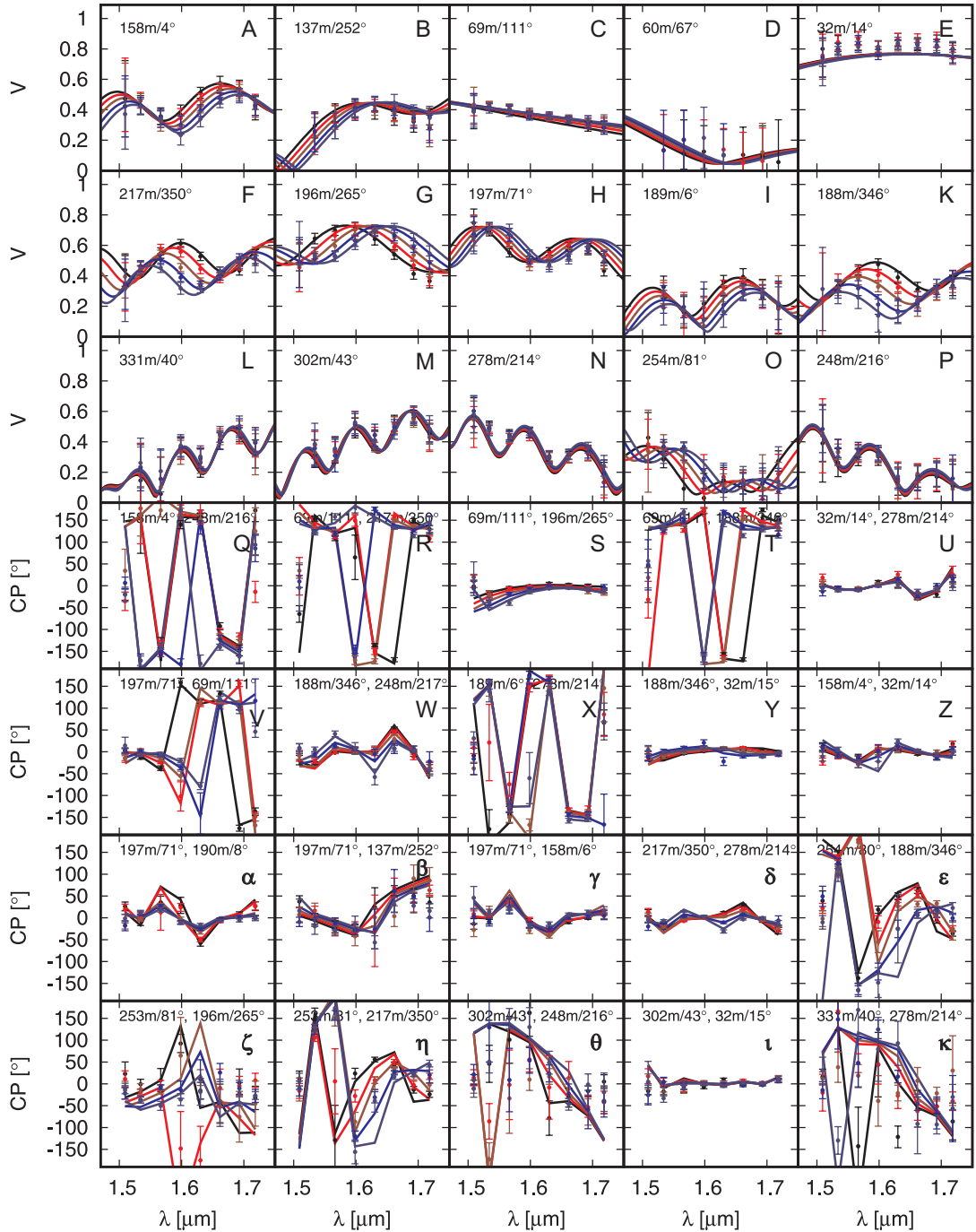

Figure S7 – **Same as Fig. S5, but for the interferometric observables recorded with CHARA/MIRC-X on 2019 August 27.** The panels show visibility, V (panel A-P) with the baseline length and PA labeled within the panel), and closure phase, CP (panel Q-$\kappa$, with the length/PA of two of the three baselines in the closed baseline triangle indicated). For each baseline/closure-phase triangle, five separate measurements are shown in different colors. The observables are over-plotted with the triple star+smooth ring model outlined in Sect. S1.2.1 (solid curves).

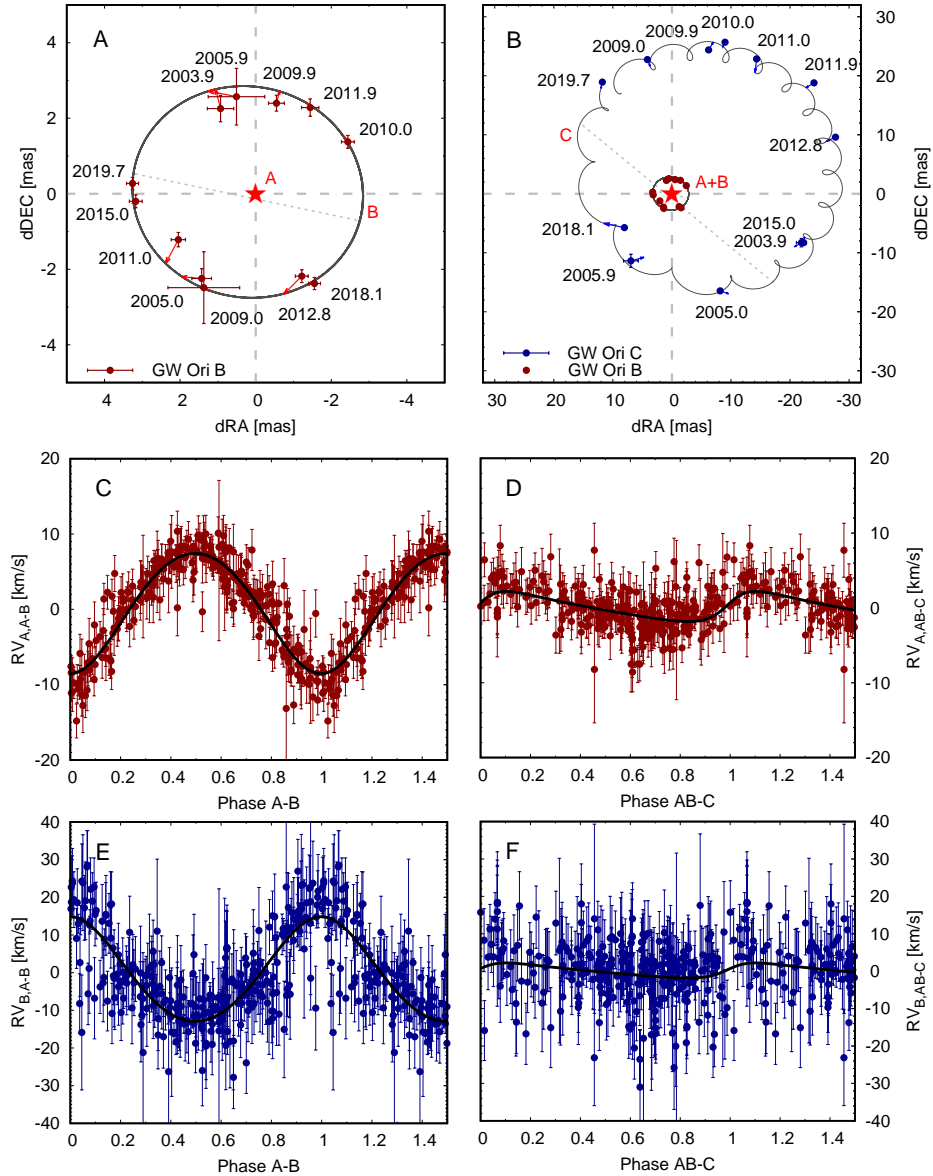

Figure S8 – **Astrometry+RV measurements on GW Ori and hierarchical triple orbit solution.** (A) and (B): Astrometry measurements of the GW Orionis inner short-period binary system A-B (panel A) and long-period system AB-C (panel B), centered on primary A. Overplotted is our best-fitting orbit (solid lines) and the line of nodes (dotted lines). The labels indicate the year an observation was taken, while the arrows mark the companion position predicted by the model for the time of the observation. (C) and (D): Radial velocity of photospheric lines associated with the primary ($RV_A$) after subtracting the motion due to the AB-C orbit (panel C) and to due to the A-B orbit (panel D), respectively (*15*). The black curve in panels C-F shows the best-fitting model that describes the motion of the A-B (panel C) and AB-C system (panel D). (E) and (F): Radial velocity of photospheric lines associated with the secondary ($RV_B$) after subtracting the motion due to the AB-C orbit (panel E) and to due to the A-B orbit (panel F) (*15*).

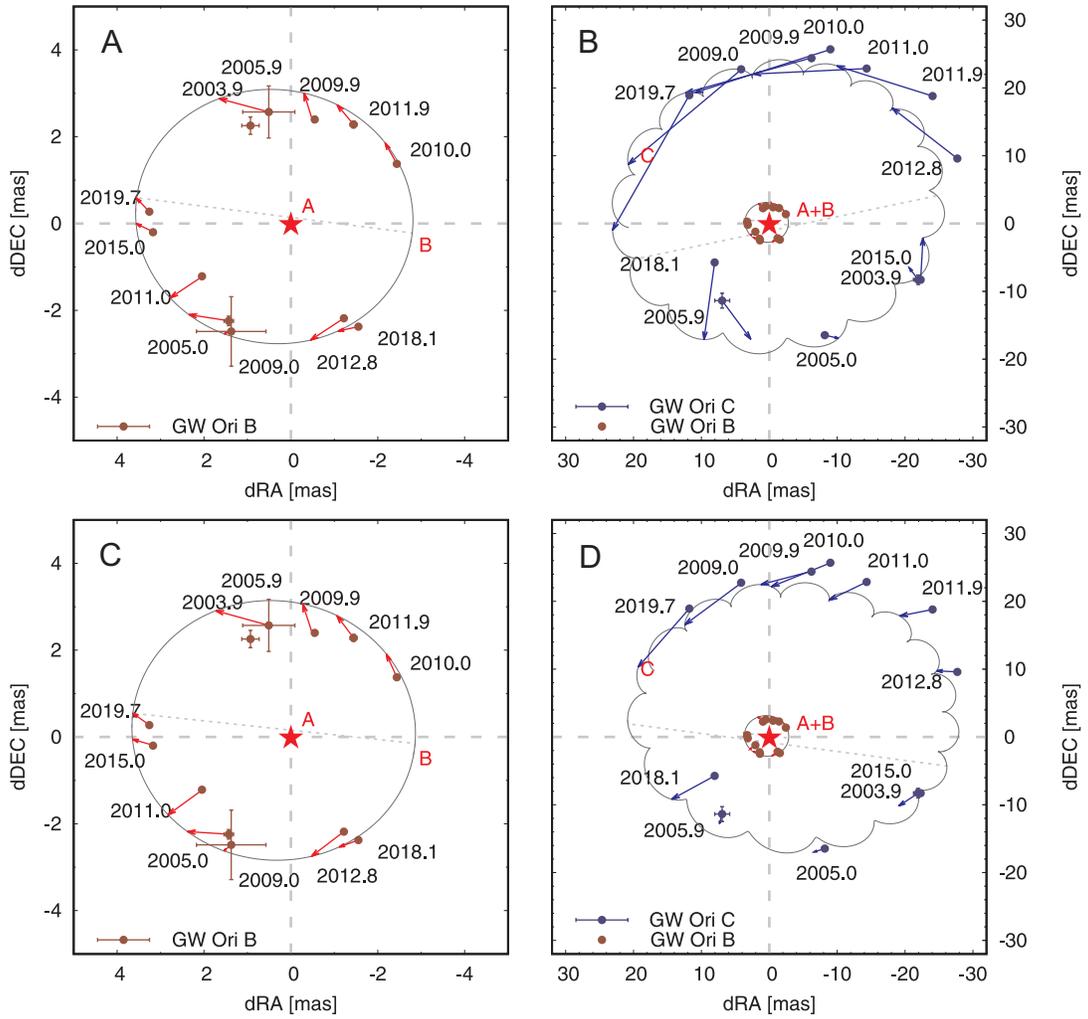

Figure S9 – **Same as Fig. S8A-B, but for two alternative sets of previously published orbital solutions (*15*).**

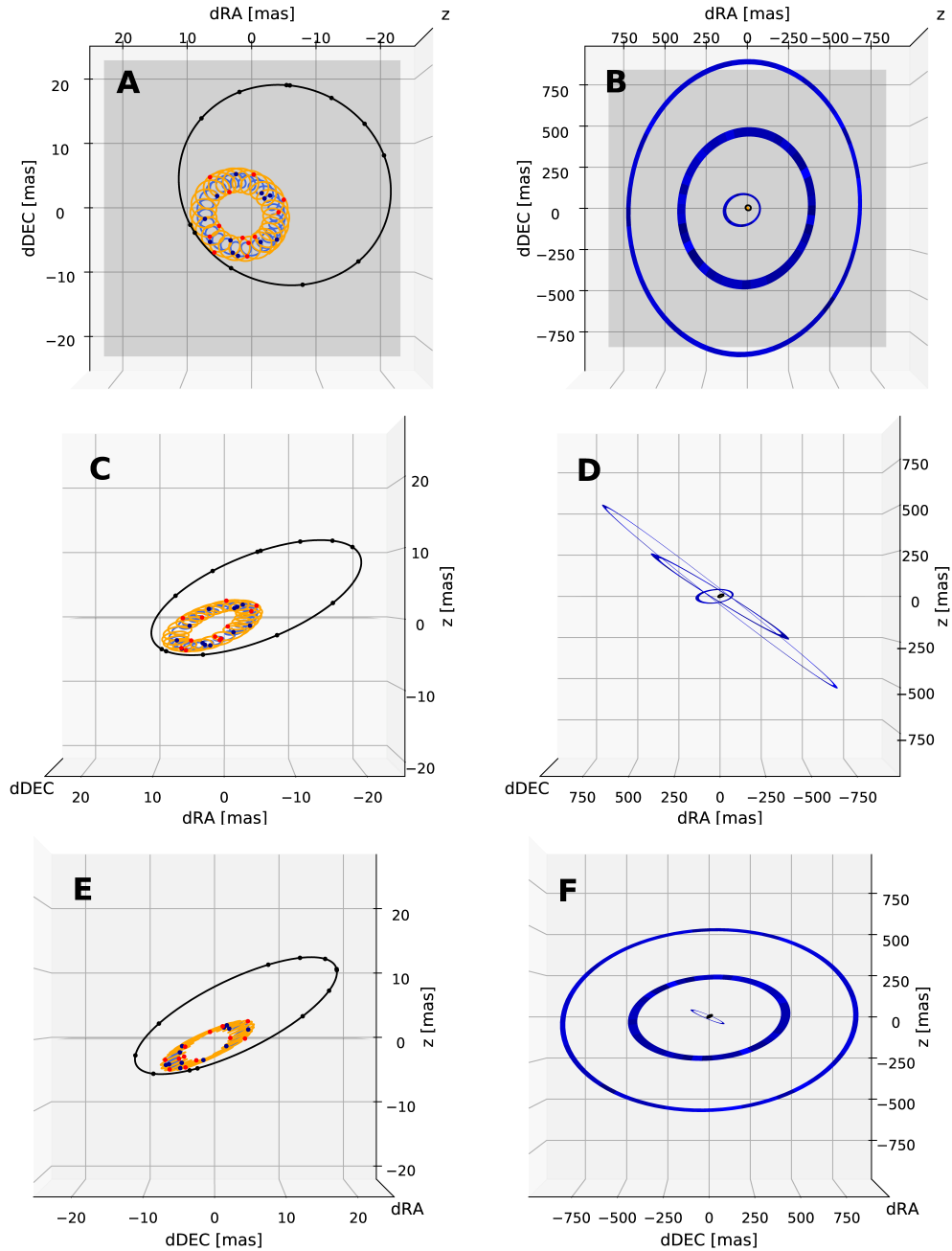

Figure S10 – **3-dimensional orientation of the stellar orbits and the brightest sub-millimeter rings R1, R2, and R3.** The black, orange, and green curves represent the orbits of the stellar components A, B, and C, while the blue rings represent the sub-millimeter rings R1, R2, and R3. The origin is the center-of-mass of the system. (A) and (B): On-sky projection; (C) and (D): View from South; (E) and (F): View from East. Positive values on the $\Delta z$ axis point towards the observer. Panels (A), (C), (E) and panels (B), (D), and (F) show different spatial scales. The points in the orbit curves indicates the position of the components at the times when our astrometric observations were recorded.

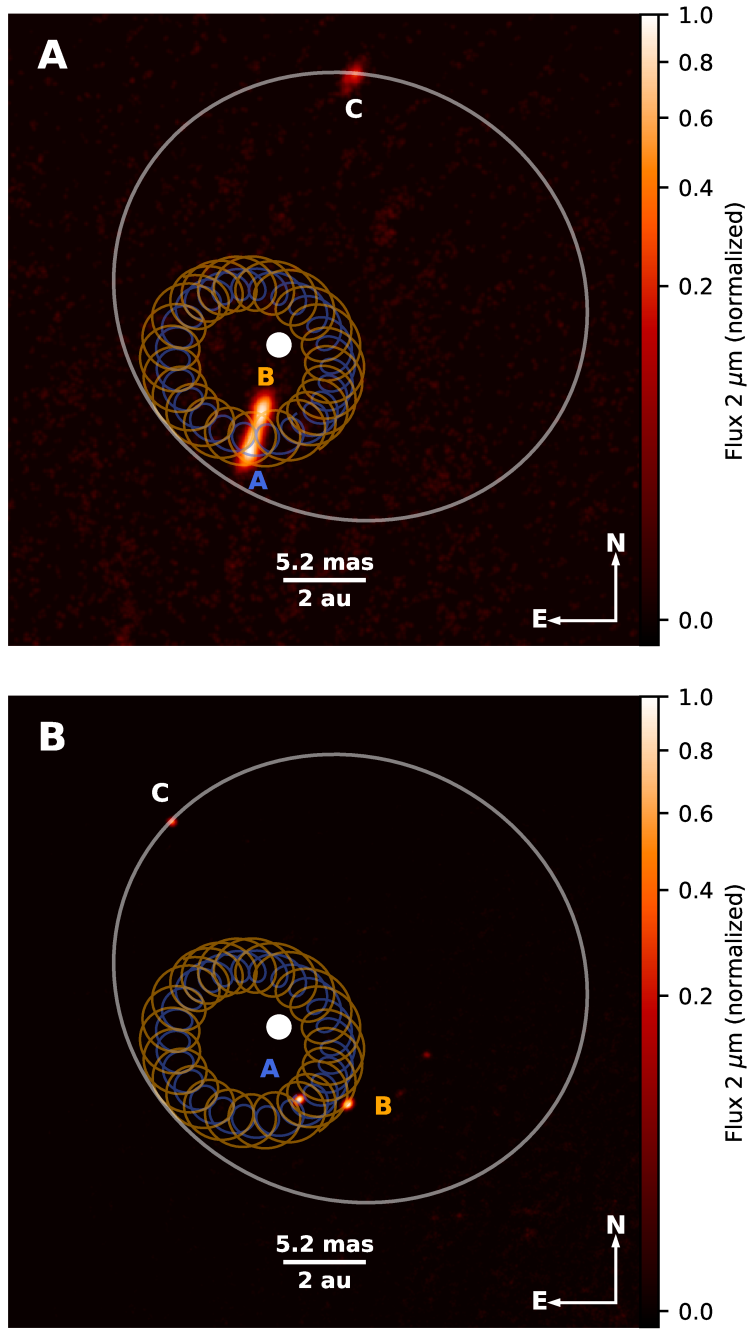

Figure S11 – **Aperture synthesis images, overplotted with the best-fitting orbital model (same colors as in Fig. S10).** The images were reconstructed from the AMBER 2009 December 4 (A) and MIRC-X 2019 August 27 data (B). The white dot marks the center-of-mass of the system.

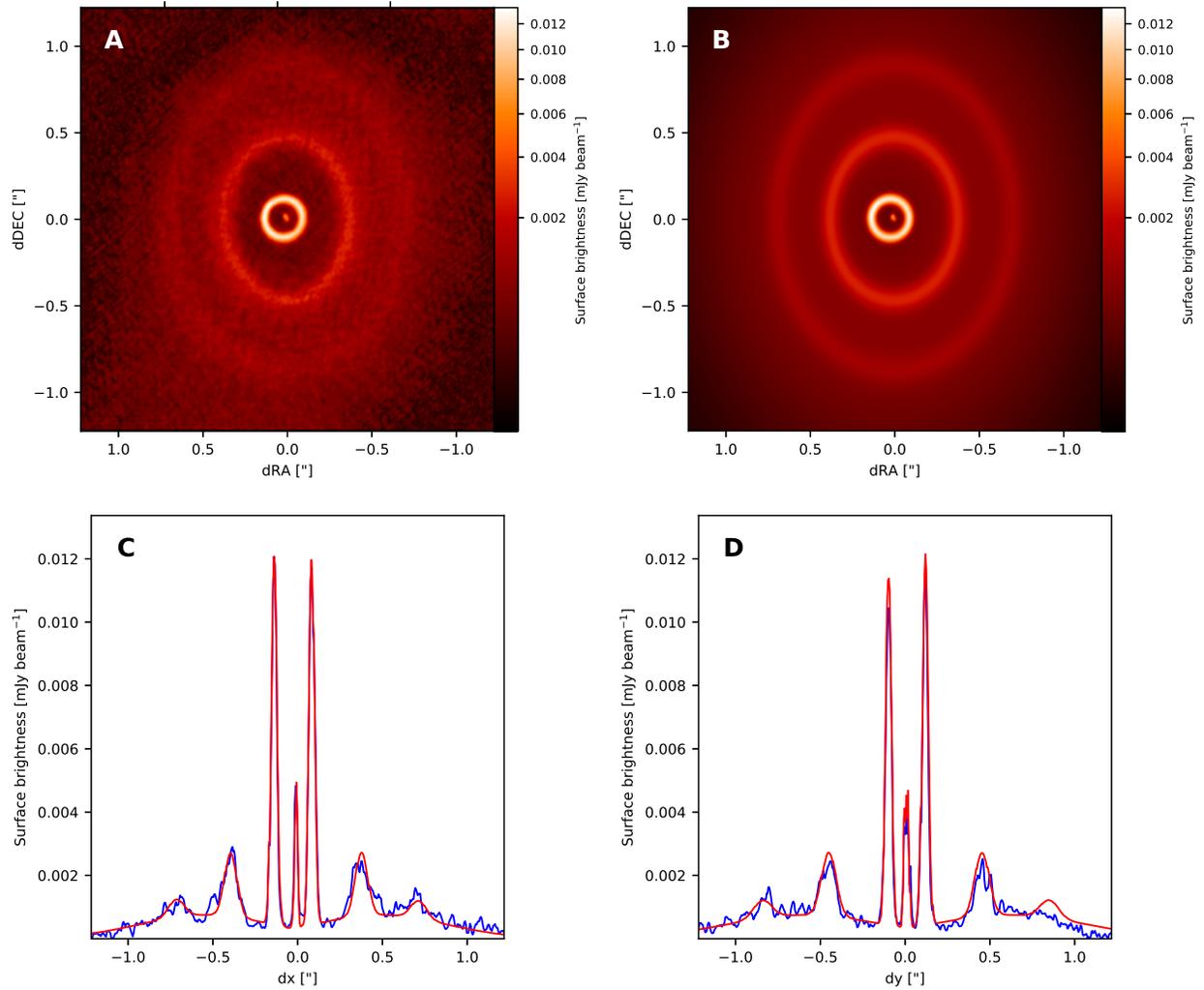

Figure S12 – **Geometric-model fit to the ALMA data.** (A) and (B): ALMA image and synthetic image (panel A) for our best-fitting model (panel B). (C) and (D): Radial intensity cuts along the disk major (panel C) and minor axis (panel D), for the ALMA image (blue line) and the model (red line).

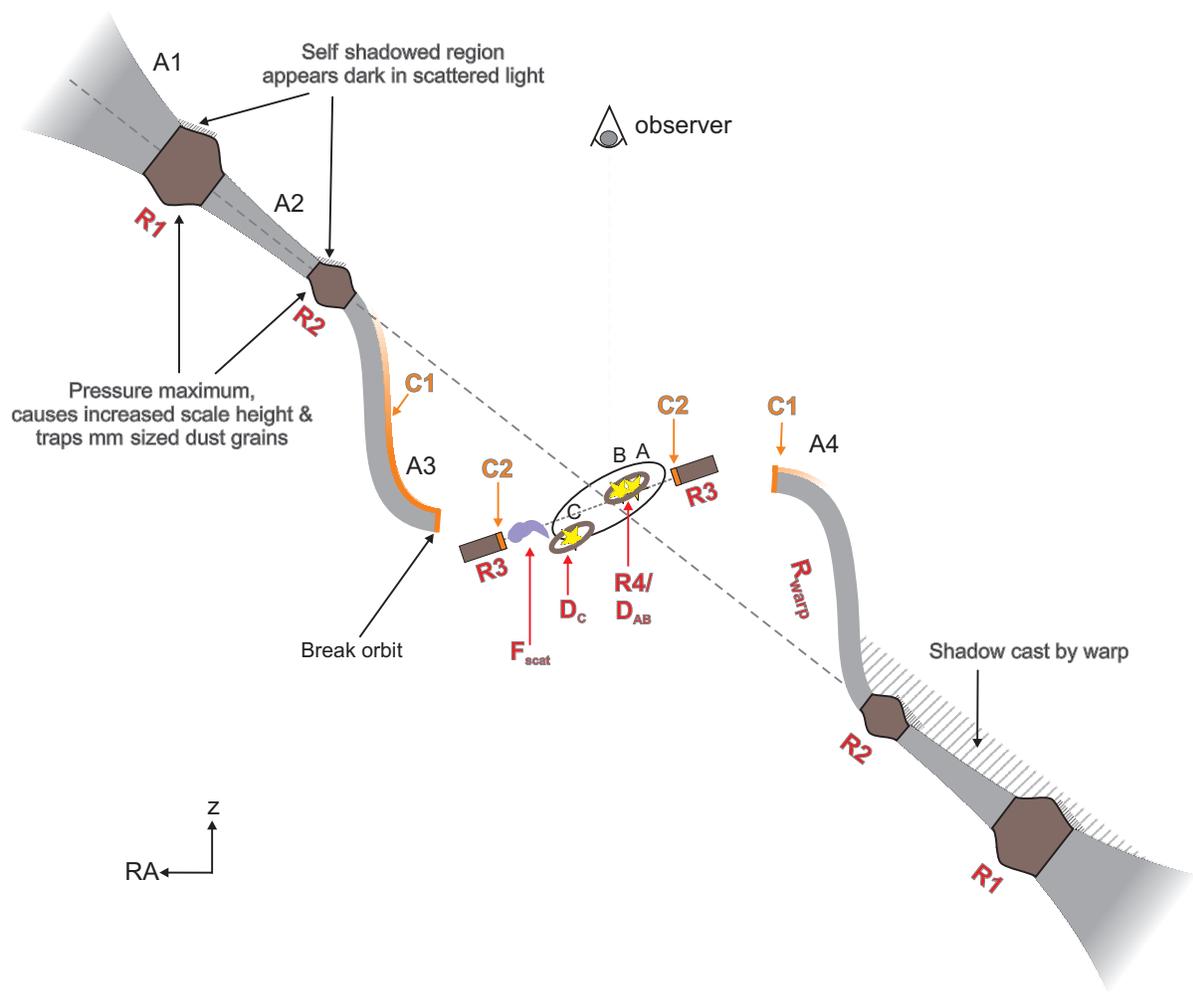

Figure S13 – **Schematic illustration of our model components and the viewing geometry**. Not to scale.

Table S1. **Observation log.** UT indicates the universal time at the beginning of the data recording, DIT the detector integration time, and NEXP the number of exposures that have been recorded

| Instrument | Spectral setting | Date (UT) | Time (UT) | DIT [s] | NEXP | Telescope(s) configuration | Calibrators |
|---|---|---|---|---|---|---|---|
| AMBER | LR-HK | 2008-12-15 | 04:30 | 0.050 | 5 | UT1-UT3-UT4 | HD 29150, HD 30912 |
| AMBER | LR-HK | 2009-12-05 | 05:21 | 0.050 | 10 | UT1-UT2-UT4 | HD 34203 |
| AMBER | LR-HK | 2009-12-06 | 06:43 | 0.021 | 10 | UT1-UT3-UT4 | HD 34203 |
| AMBER | LR-HK | 2009-12-06 | 07:49 | 0.021 | 8 | UT1-UT3-UT4 | HD 34203 |
| AMBER | LR-HK | 2009-12-06 | 08:51 | 0.021 | 8 | UT1-UT3-UT4 | HD 34203 |
| AMBER | LR-HK | 2009-12-31 | 02:20 | 0.021 | 12 | UT1-UT2-UT4 | HD 37128 |
| AMBER | LR-HK | 2009-12-31 | 03:05 | 0.021 | 6 | UT1-UT2-UT4 | HD 37128 |
| AMBER | LR-HK | 2010-01-01 | 03:50 | 0.021 | 10 | UT1-UT3-UT4 | HD 34203 |
| AMBER | LR-HK | 2010-01-01 | 04:24 | 0.021 | 6 | UT1-UT3-UT4 | HD 34203 |
| AMBER | LR-HK | 2010-01-01 | 07:12 | 0.021 | 12 | UT2-UT3-UT4 | HD 42807 |
| AMBER | LR-HK | 2010-12-17 | 02:55 | 0.050 | 10 | UT2-UT3-UT4 | HD 34203 |
| AMBER | LR-HK | 2010-12-17 | 03:41 | 0.050 | 15 | UT2-UT3-UT4 | HD 41794 |
| AMBER | LR-HK | 2010-12-18 | 07:10 | 0.027 | 5 | UT2-UT3-UT4 | HD 34203 |
| AMBER | LR-HK | 2011-12-09 | 04:02 | 0.026 | 10 | UT1-UT3-UT4 | HD 34203 |
| AMBER | LR-HK | 2011-12-09 | 05:27 | 0.026 | 10 | UT1-UT3-UT4 | HD 34203 |
| AMBER | LR-HK | 2012-10-29 | 05:20 | 0.026 | 6 | UT1-UT2-UT4 | HD 22781 |
| AMBER | LR-HK | 2012-10-29 | 06:38 | 0.026 | 10 | UT1-UT2-UT4 | HD 34203 |
| AMBER | LR-HK | 2012-10-29 | 09:15 | 0.026 | 5 | UT1-UT2-UT4 | HD 28462 |
| AMBER | LR-HK | 2015-01-01 | 02:17 | 0.026 | 5 | UT1-UT3-UT4 | HD 37128 |
| MIRC-X | LR-H | 2019-08-27 | 12:06 | 0.0028 | 5 | S1-S2-E1-E2-W1-W2 | HD 240579 |
| Gravity* | MR | 2017-10-25 | 06:14 | 30 | 6 | A0-B2-C1-D0 | HD 244179, HD 38117 |
| Gravity* | MR | 2017-10-25 | 07:43 | 30 | 6 | A0-B2-C1-D0 | HD 244179, HD 37926 |
| Gravity | MR | 2018-02-06 | 01:30 | 30 | 4 | A0-G1-J2-J3 | HD 38117 |
| GPI | J-coron-pol | 2018-01-04 | 02:33 | 29 | 2 × 35 | Gemini-S | – |
| GPI | H-coron-pol | 2018-01-04 | 03:59 | 29 | 2 × 35 | Gemini-S | – |
| SPHERE/ZIMPOL | I'-band, SlowPol | 2018-10-15 | 06:58 | 8 | 6 × 4 | UT3 | – |
| SPHERE/IRDIS | H-band, SlowPol | 2018-11-16 | 06:07 | 16 | 9 × 6 | UT3 | – |
| ALMA** | continuum 1.3 mm | 2015-05-14 | 20:42 | 1170 | | C34-3/4 | QSO B0507+179, QSO B0507+179 |
| ALMA | continuum 1.3 mm | 2019-07-05 | 13:47 | 2782 | | C43-9/10 | QSO J0530+13, QSO B0507+179 |

*not included in the astrometry model, as the resolution achieved in the K-band on the compact VLTI configuration was not sufficient to resolve the inner binary (A-B).

**previously-published data set (*15*).

Table S2.  **Information on the interferometric calibrator stars listed in Table S1.**

| Star | $H$ [magnitude] | $K$ [magnitude] | Spectral Type | $d_{\rm UD}$ [mas] | Reference |
|---|---|---|---|---|---|
| HD 22781 | 6.69 | 6.61 | K0 | $0.228 \pm 0.016$ | (84, 85) |
| HD 28462 | 7.21 | 7.14 | K1 | $0.179 \pm 0.004$ | (84, 85) |
| HD 29150 | 6.08 | 5.99 | G5 | $0.285 \pm 0.020$ | (84, 85) |
| HD 30912 | 5.07 | 4.98 | F2IV | $0.407 \pm 0.028$ | (84, 85) |
| HD 34203 | 5.47 | 5.46 | A0V | $0.228 \pm 0.016$ | (84, 85) |
| HD 37128 | 2.41 | 2.27 | B0Iab | $0.670 \pm 0.040$ | (84, 85) |
| HD 37926 | 5.60 | 5.44 | K0 | $0.393 \pm 0.009$ | (84, 85) |
| HD 38117 | 5.12 | 4.91 | K0 | $0.494 \pm 0.016$ | (84, 85) |
| HD 41794 | 6.19 | 6.13 | A5 | $0.213 \pm 0.015$ | (84, 85) |
| HD 42807 | 5.01 | 4.85 | G2V | $0.486 \pm 0.034$ | (84, 85) |
| HD 240579 | 6.37 | 6.15 | K0 | $0.293 \pm 0.007$ | (84, 85) |
| HD 244179 | 6.02 | 5.87 | K0 | $0.327 \pm 0.008$ | (84, 85) |

Table S3.  **Adopted system parameters for GW Orionis.** We follow earlier studies (e.g. (*15*)) by adopting the distance estimate from (*11*). Their distance estimate reports smaller measurement uncertainties than other studies and is compatible with the Gaia DR2 (*66, 67*) value of $398 \pm 10$ pc and avoid potentially biased due to astrometric motion of the triple system.

|  | Symbol | Unit | Value | Reference |
|---|---|---|---|---|
| Distance[a] | $d$ | [pc] | $388 \pm 5$ | (*11*) |
| Effective temperature | $T_{\rm eff}$ | [K] | 5500 | (*45*) |
| Bolometric luminosity | $L_{\rm sun}$ | [$L_{\rm sun}$] | $48 \pm 10$ | (*24*) |
| Extinction | $A_V$ | [mag] | $1.5 \pm 0.1$ | (*24*) |
| Disk total mass (gas+dust) | $M_{\rm disk}$ | [$M_{\rm sun}$] | 0.12 | (*45*) |

Table S4. **Triple star astrometry derived from near-infrared interferometric data.** $F_B/F_A$ and $F_C/F_A$ denote the flux ratio between the B or C and A component, $\rho_{AB}$ and $\rho_{AC}$ the separation between the A and B or C component, and $\theta_{AB}$ and $\theta_{AC}$ are the PA of the B or C component with respect to the A component. $F_{\text{ext}}/F_{\text{tot}}$ denotes the fraction of the flux that is contributed by the extended emission component to the total flux in our model.

| Epoch | $F_B/F_A$ | $\rho_{AB}$ [mas] | $\theta_{AB}$ [°] | $F_C/F_A$ | $\rho_{AC}$ [mas] | $\theta_{AC}$ [°] | $F_{\text{ext}}/F_{\text{tot}}$ | $\chi^2_{\text{red}}$ |
|---|---|---|---|---|---|---|---|---|
| 2003.913 | $0.88 \pm 0.19$ | $2.5 \pm 0.2$ | $22.4 \pm 6$ | $0.22 \pm 0.03$ | $23.5 \pm 0.7$ | $250 \pm 2$ | $0.16^*$ | 0.3 |
| 2004.957 | $1.10 \pm 0.25$ | $2.8 \pm 0.8$ | $151 \pm 3$ | $0.36 \pm 0.06$ | $18.4 \pm 0.2$ | $206 \pm 1$ | $0.16^*$ | 2.1 |
| 2005.890 | $0.65 \pm 0.44$ | $2.6 \pm 0.6$ | $11 \pm 2$ | $0.31 \pm 0.17$ | $13.3 \pm 1.1$ | $148 \pm 6$ | $0.16^*$ | 0.2 |
| 2008.956 | $0.66 \pm 0.06$ | $2.66 \pm 0.11$ | $147.5 \pm 0.9$ | $0.19 \pm 0.02$ | $23.12 \pm 0.18$ | $10.3 \pm 0.5$ | $0.16^*$ | 0.7 |
| 2009.927 | $0.97 \pm 0.16$ | $2.46 \pm 0.06$ | $347.1 \pm 0.5$ | $0.21 \pm 0.02$ | $25.16 \pm 0.21$ | $345.7 \pm 0.1$ | $0.16 \pm 0.02$ | 0.5 |
| 2009.995 | $0.74 \pm 0.02$ | $2.80 \pm 0.02$ | $299.4 \pm 0.3$ | $0.20 \pm 0.02$ | $27.22 \pm 0.08$ | $340.7 \pm 0.1$ | $0.16^*$ | 0.2 |
| 2010.960 | $1.08 \pm 0.10$ | $2.38 \pm 0.04$ | $120.7 \pm 1.3$ | $0.23 \pm 0.01$ | $27.00 \pm 0.25$ | $328.0 \pm 0.3$ | $0.16^*$ | 1.1 |
| 2011.937 | $0.71 \pm 0.12$ | $2.70 \pm 0.08$ | $327.7 \pm 0.6$ | $0.26 \pm 0.03$ | $30.53 \pm 0.19$ | $308.0 \pm 0.2$ | $0.16^*$ | 0.9 |
| 2012.825 | $1.12 \pm 0.03$ | $2.50 \pm 0.02$ | $209.2 \pm 0.6$ | $0.32 \pm 0.01$ | $29.3 \pm 0.1$ | $289.1 \pm 0.2$ | $0.16^*$ | 0.9 |
| 2015.000 | $0.64 \pm 0.03$ | $3.18 \pm 0.02$ | $93.6 \pm 0.4$ | $0.15 \pm 0.01$ | $23.8 \pm 0.2$ | $249.6 \pm 0.3$ | $0.16^*$ | 0.5 |
| 2018.099 | $0.83 \pm 0.01$ | $2.84 \pm 0.01$ | $213.2 \pm 0.2$ | $0.47 \pm 0.01$ | $9.9 \pm 0.1$ | $125.5 \pm 0.1$ | $0.16^*$ | 6.2 |
| 2019.652 | $0.96 \pm 0.01$ | $3.27 \pm 0.01$ | $85.2 \pm 0.1$ | $0.37 \pm 0.01$ | $22.3 \pm 0.01$ | $31.9 \pm 0.1$ | $0.16^*$ | 4.8 |

$^*$fixed in the modeling process.

Table S5. **Orbital elements of the GW Ori triple star system derived from astrometry.** $P$ denotes the orbital period, $e$ the eccentricity, $T_0$ the time of periastron passage, $\gamma$ the velocity of the system center-of-mass, $i$ the orbital inclination, $a$ the orbital semi-major axis, and $K_A$, $K_B$, $K_C$ the semi-amplitude in radial velocity of the stellar components. $q = M_B/M_A$ is the mass ratio of the A and B component. $\Omega$ gives the longitude of the ascending node (i.e. the node where the motion of the secondary is directed away from the Sun) and $\omega$ is the longitude of the periastron, measured from the ascending node of the secondary. JD is the Julian Day Number.

| Param. | | Ref. (15) solution #1 | | Ref. (15) solution #2 | | This study | |
|---|---|---|---|---|---|---|---|
| | | Orbit $A-B$ | Orbit $(AB)-C$ | Orbit $A-B$ | Orbit $(AB)-C$ | Orbit $A-B$ | Orbit $(AB)-C$ |
| $P$ | [d] | $241.50 \pm 0.05$ | $4246 \pm 66$ | $241.49 \pm 0.04$ | $4203 \pm 60$ | $241.619 \pm 0.05$ | $4216.8 \pm 4.6$ |
| $e$ | | $0.13 \pm 0.01$ | $0.13 \pm 0.07$ | $0.13 \pm 0.01$ | $0.25 \pm 0.08$ | $0.069 \pm 0.009$ | $0.379 \pm 0.003$ |
| $\omega$ | [°] | $17 \pm 7$ | $130 \pm 21$ | $16 \pm 6$ | $130 \pm 12$ | $1 \pm 7$ | $105 \pm 1$ |
| $T_0$ | JD | $2456682 \pm 4$ | $2453911 \pm 260$ | $2456681 \pm 4$ | $2453878 \pm 130$ | $2456674.8 \pm 4.7$ | $2453859.6 \pm 4.8$ |
| $\gamma$ | [km s$^{-1}$] | $+28.33 \pm 0.18$ | – | $+28.29 \pm 0.19$ | – | $+27.96 \pm 0.12$ | – |
| $q$ | | $0.60 \pm 0.02$ | – | $0.60 \pm 0.02$ | – | $0.60 \pm 0.02$ | – |
| $i$ | [°] | $157 \pm 1$ | $150 \pm 7$ | $157 \pm 1$ | $144 \pm 9$ | $156 \pm 1$ | $149.6 \pm 0.7$ |
| $\Omega$ | [°] | $263 \pm 13$ | $282 \pm 9$ | $264 \pm 13$ | $263 \pm 10$ | $258.2 \pm 1.3$ | $230.9 \pm 1.1$ |
| $a$ | [mas] | $3.2 \pm 0.2$ | $23.7 \pm 0.8$ | $3.3 \pm 0.2$ | $23.6 \pm 0.9$ | $3.08 \pm 0.1$ | $22.9 \pm 0.1$ |
| $a$ | [au] | $1.28 \pm 0.05$ | $9.43 \pm 0.33$ | $1.31 \pm 0.05$ | $9.39 \pm 0.36$ | $1.20 \pm 0.04$ | $8.89 \pm 0.04$ |
| $K_A$ | [km s$^{-1}$] | $8.34 \pm 0.15$ | | $8.36 \pm 0.15$ | | $7.98 \pm 0.16$ | |
| $K_B$ | [km s$^{-1}$] | | | | | $13.88 \pm 0.38$ | |
| $K_C$ | [km s$^{-1}$] | | $2.38 \pm 0.23$ | | $2.50 \pm 0.24$ | | $2.01 \pm 0.20$ |
| $M_{\text{tot}}$ | [M$_\odot$] | $5.7 \pm 0.7$ | | $6.1 \pm 0.9$ | | $3.90 \pm 0.40$ | $5.26 \pm 0.22$ |
| $M_A$ | [M$_\odot$] | $2.80^{+0.36}_{-0.31}$ | | – | | $2.47 \pm 0.33$ | |
| $M_B$ | [M$_\odot$] | $1.68^{+0.21}_{-0.18}$ | | – | | $1.43 \pm 0.18$ | |
| $M_C$ | [M$_\odot$] | $1.15^{+0.40}_{-0.23}$ | | – | | $1.36 \pm 0.28$ | |

Table S6.  **Disk model-fitting results for the 1.3 mm ALMA data.** For all rings we identify the southern node as the ascending node, which is consistent with the rings being in retrograde rotation (i.e. following the rotation direction of the stellar orbits) and the Northern part of the rings receeding from the observer, as derived from the CO rotation measurements (*15, 45*). The semi-major axis estimate $b_{R3}$ and inclination estimate $i_{R3,e=0}$ assume that the ring has no intrinsic eccentricity, i.e. that it is intrinsically centro-symmetric. As before, we define inclination 0° as face-on. PAs are measured East of North and along the disk minor axis.

| Model parameter | Symbol | Unit | Best-fit value |
|---|---|---|---|
| **Dust near location of primary/secondary $D_{AB}$ (1.3mm)** | | | |
| Flux density | $F_{\nu,AB}$ | [mJy] | 0.06 |
| HWHM | $\Theta_{AB}$ | [mas] | $\lesssim 4$ |
| Derived dust mass | $M_{D,AB}$ | [$M_{Earth}$] | $0.02 \pm 0.01$ |
| **Dust near location of tertiary $D_C$ (1.3mm)** | | | |
| Flux density | $F_{\nu,C}$ | [mJy] | 0.18 |
| HWHM | $\Theta_C$ | [mas] | $\lesssim 4$ |
| Offset dRA | $\Delta\alpha_C$ | [mas] | $10.3 \pm 2$ |
| Offset dDec | $\Delta\delta_C$ | [mas] | $-13.0 \pm 2$ |
| Derived dust mass | $M_{D,C}$ | [$M_{Earth}$] | $0.05 \pm 0.02$ |
| **Ring R3 (1.3mm thermal dust imaging)** | | | |
| Flux density | $F_{\nu,R3}$ | [mJy] | $14 \pm 2$ |
| Semi-minor axis ($e=0$, fixed) | $b_{R3}$ | [mas] | $112 \pm 3$ (=$43.3 \pm 1.1$ au) |
| Inclination ($e=0$, fixed) | $i_{R3,e=0}$ | [°] | $179.9 \pm 1.0$ |
| HWHM | $\Theta_{R3}$ | [mas] | $16.5 \pm 4.0$ |
| Offset dRA | $\Delta\alpha_{R3}$ | [mas] | $26.2 \pm 0.25$ |
| Offset dDec | $\Delta\delta_{R3}$ | [mas] | $-5.4 \pm 0.15$ |
| Asymmetry, PA | $\theta_{R3asym}$ | [°] | $0 \pm 9$ |
| Asymmetry, amplitude | $a_{R3asym}$ | | $0.07 \pm 0.04$ |
| Asymmetry, stretch-factor | $\gamma_{R3asym}$ | | $0.84 \pm 0.03$ |
| Derived dust mass | $M_{R3}$ | [$M_{Earth}$] | $30 \pm 4$ |
| **(including 1.6 µm/0.8 µm scattered light imaging constraints)** | | | |
| Inclination ($e$ free) | $i_{R3}$ | [°] | $155 \pm 16$ |
| Eccentricity | $e_{R3}$ | | $0.3 \pm 0.1$ |
| Ascending node | $\Omega_{R3}$ | [°] | $285 \pm 30$ |
| Semi-major axis ($e$ free) * | $a_{R3}$ | [mas] | 122 (=47 au) |
| **Ring R2 (1.3mm thermal dust imaging)** | | | |
| Flux density | $F_{\nu,R2}$ | [mJy] | $22 \pm 2$ |
| Radius | $r_{R2}$ | [mas] | $470 \pm 32$ (=$182 \pm 12$ au) |
| Inclination | $i_{R2}$ | [°] | $143 \pm 1$ |
| HWHM | $\Theta_{R2}$ | [mas] | $64.1 \pm 15$ |
| Ascending node | $\Omega_{R2}$ | [°] | $180 \pm 4$ |
| Derived dust mass | $M_{R2}$ | [$M_{Earth}$] | $168 \pm 25$ |
| **Ring R1 (1.3mm thermal dust imaging)** | | | |
| Flux density | $F_{\nu,R1}$ | [mJy] | $15 \pm 2$ |
| Radius | $r_{R1}$ | [mas] | $906 \pm 31$ (=$351 \pm 12$ au) |
| Inclination | $i_{R1}$ | [°] | $142 \pm 1$ |
| HWHM | $\Theta_{R1}$ | [mas] | $95 \pm 20$ |
| Ascending node | $\Omega_{R1}$ | [°] | $180 \pm 8$ |
| Derived dust mass | $M_{R1}$ | [$M_{Earth}$] | $153 \pm 23$ |
| **Extended disk $R_{disk}$ (1.3mm thermal dust imaging)** | | | |
| Flux density | $F_{\nu,Rdisk}$ | [mJy] | $109 \pm 11$ |
| Radius | $r_{Rdisk}$ | [mas] | $625 \pm 70$ (=$242 \pm 27$ au) |
| HWHM | $\Theta_{Rdisk}$ | [mas] | $556 \pm 113$ |
| Derived dust mass | $M_{Rdisk}$ | [$M_{Earth}$] | $225 \pm 35$ |

*Not fitted, but determined to match the on-sky projected shape of R3 based on the fitted values for $i_{R3}$, $e_{R3}$, and $\Omega_{R3}$.